\documentclass{ifcolog}

\usepackage{amsmath}

\usepackage{amsfonts}
\usepackage{dsfont}

\usepackage{amssymb}

\usepackage{amsmath}

\usepackage{amsthm}

\usepackage{graphicx}

\usepackage{epstopdf}

\usepackage[]{hyperref}

\usepackage{float}

\usepackage{subcaption}
\usepackage{array}
\usepackage{makecell}

\usepackage{epsfig}

\usepackage{array,longtable}

\usepackage{hyperref}

\usepackage{mdframed}

\usepackage{lipsum}
\newtheorem{definition}{Definition}[section]

\newtheorem{lemma}{Lemma}[section]
\newtheorem{subgoal}{Subgoal}[section]

\newtheorem{theorem}{Theorem}[section]

\usepackage{url}

\PassOptionsToPackage{hyphens}{url}\usepackage{hyperref}

\title{Formalization of Lerch's Theorem using HOL Light}

\titlerunning{Formalization of Lerch's Theorem using HOL Light}

\addauthor[adnan.rashid@seecs.nust.edu.pk]{Adnan Rashid}{School of Electrical Engineering and Computer Science (SEECS) \\ National University of Sciences and Technology (NUST), Islamabad, Pakistan}

\addauthor[osman.hasan@seecs.nust.edu.pk]
  {Osman Hasan}
  {School of Electrical Engineering and Computer Science (SEECS) \\ National University of Sciences and Technology (NUST), Islamabad, Pakistan}

\authorrunning{A. Rashid and O. Hasan}

\titlethanks{}

\begin{document}
\maketitle
\begin{abstract}
The Laplace transform is an algebraic method that is widely used for analyzing physical systems by either solving the differential equations modeling their dynamics or by evaluating their transfer function. The dynamics of the given system are firstly modeled using differential equations and then Laplace transform is applied to convert these differential equations to their equivalent algebraic equations. These equations can further be simplified to either obtain the transfer function of the system or to find out the solution of the differential equations in frequency domain. Next, the uniqueness of the Laplace transform provides the solution of these differential equations in the time domain. The traditional Laplace transform based analysis techniques, i.e., paper-and-pencil proofs and computer simulation methods are error-prone due to their inherent limitations and thus are not suitable for the analysis of the systems. Higher-order-logic theorem proving can overcome these limitations of these techniques and can ascertain accurate analysis of the systems. In this paper, we extend our higher-order logic formalization of the Laplace transform, which includes the formal definition of the Laplace transform and verification of its various classical properties. One of the main contributions of the paper is the formalization of Lerch's theorem, which describes the uniqueness of the Laplace transform and thus plays a vital role in solving linear differential equations in the frequency domain. For illustration, we present the formal analysis of a $4$-$\pi$ soft error crosstalk model, which is widely used in nanometer technologies, such as, Integrated Circuits (ICs).
\end{abstract}


\section{Introduction}\label{SEC:Introduction}

The engineering and physical systems exhibiting the continuous-time dynamical behaviour are mathematically modeled using differential equations, which need to be solved to judge system characteristics. Laplace transform method allows us to solve these differential equations or evaluate the transfer function of the signals in these systems using algebraic techniques and thus is very commonly used in system analysis. Taking the Laplace transform of differential equations allows us to convert the time-varying functions involved in these differential equations to their corresponding $s$-domain representations, i.e., the integral and differential operators in time domain are converted to their equivalent multiplication and division operators in the $s$-domain, where $s$ represents the angular frequency. These algebraic equations can then be further simplified to either obtain the transfer function of the system or solution of the differential equations in frequency domain. In the last step, the uniqueness of the Laplace transform is used to obtain the solution of these differential equations in time domain.

Traditionally, the Laplace transform is used for analyzing the engineering and physical systems using paper-and-pencil proofs, numerical methods and symbolic techniques. However, these analysis techniques cannot ascertain accuracy due to their inherent limitations, like human-error proneness, discretization and numerical errors.
For example, the Laplace transform based analysis provided by the computer algebra systems, like Mathematica and Maple, and Symbolic Math Toolbox of Matlab use the algorithms that consider the improper integral involved in the definition of the Laplace transform as the continuous analog of the power series, i.e., the integral is discretized to summation and the complex exponentials are sampled~\cite{taqdees2013formalization}.
Given the wide-spread usage of these systems in many safety-critical domains, such as medicine, military and transportation, accurate transform method based analysis has become a dire need. With the same motivation, the Laplace transform has been formalized in the HOL Light theorem prover and it has been successfully used for formally analyzing the Linear Transfer Converter (LTC) circuit~\cite{taqdees2013formalization}, Sallen key low-pass filters~\cite{taqdees2017tflac}, Unmanned Free-swimming Submersible (UFSS) vehicle~\cite{rashid2017formal} and platoon of the automated vehicles~\cite{adnan2018SEFM}. Similarly, the Fourier transform~\cite{bracewell1965fourier} has also been formalized in the same theorem prover and has been used for formally analyzing an Automobile Suspension System (ASS)~\cite{rashid2016formalization}, audio equalizer~\cite{adnanJSC2017} and MEMs accelerometer~\cite{adnanJSC2017}. However, both of these formalizations can only be used for the frequency domain analysis. In order to relate this frequency-domain analysis to the corresponding linear differential equations in the time domain, we need the uniqueness of the Laplace and Fourier transforms and Lerch's theorem fulfills this requirement for the former. However, to the best of our knowledge, Lerch's theorem has not been formally verified in a theorem prover so far. We overcome this limitation in this paper with the motivation that the verification of Lerch's theorem along with the existing formalization of the Laplace transform~\cite{taqdees2013formalization,rashid2017formal} would facilitate the formal reasoning about the time domain solutions of differential equations in the sound core of a theorem prover and thus the formal analysis of many engineering systems~\cite{yang2013mathematical}.
For this purpose, we extend our formalization of the Laplace transform in higher-order logic~\cite{taqdees2013formalization,rashid2017formal}, which includes the formal definition of the Laplace transform and verification of it various classical properties. We present a new definition of the Laplace transform~\cite{rashid2017formal,rashid2017formalization}, which is based on the notion of sets. Moreover, we formally verify its various properties, which include time scaling, time shifting, modulation, Laplace transform of $n$-order differential equation and transfer function of a generic $n$-order system~\cite{rashid2017formal}, in addition to the properties, which were verified using the older definition namely linearity, frequency shifting, and differentiation and integration in time domain~\cite{taqdees2013formalization}.


Lerch's theorem~\cite{cohen2007numerical,orloff_uniqueness} provides the uniqueness for the Laplace transform and thus allows to evaluate the solution of differential equations using the Laplace transform in the frequency domain~\cite{podlubny1997laplace}.
Mathematically, if


\begin{equation}\label{EQ:laplace_transform_f}
\mathcal{L} [f (t)] = F(s) = \int_0^{\infty} {f(t)e^{-s t} dt}, \ \ \ \ Re\ s \geq \gamma
\end{equation}

\noindent is satisfied by a continuous function $f$, then there is no other continuous function other than $f$ that satisfies Equation~(\ref{EQ:laplace_transform_f}). The complex term $\mathcal{L} [f (t)] = F(s)$ in the above equation represents the Laplace transform of the time varying function $f$.
The above statement can alternatively be interpreted by assuming that there is another continuous function $g$, which satisfies the following condition:


\begin{equation}\label{EQ:laplace_transform_g}
\mathcal{L} [g (t)] = G(s) = \int_0^{\infty} {g(t)e^{-s t} dt}, \ \ \ \ Re\ s \geq \gamma
\end{equation}

\noindent and if $\mathcal{L} [f (t)] = \mathcal{L} [g (t)]$, then both of the functions $f$ and $g$ are the same, i.e., $f(t) = g(t)$ in $0 \le t$ ~\cite{cohen2007numerical,beerends2003fourier}.

We found a couple of paper-and-pencil proofs of Lerch's theorem~\cite{cohen2007numerical,orloff_uniqueness} in literature and both are mainly based on the following lemma.


Let $\phi:\mathds{R} \rightarrow \mathds{R}$ be a continuous function on [0, 1] and

\begin{equation}\label{EQ:integral_lemma}
\int_0^1 {x^{n} \phi(x) dx} = 0, \ \ \ \ \ for \ \ n = 0, 1, 2, ...
\end{equation}

Then

\begin{equation}\label{EQ:conclusion_lemma}
\phi(x) = 0, \ \ \ \ in \ \  0 \leq x \leq 1
\end{equation}

In both the cases, the authors adopt different strategies for the proof of the above lemma. Cohen~\cite{cohen2007numerical} considers splitting the region of integration, i.e., interval [0, 1] into three regions, namely, [0, $a$], [$a$, $b$] and [$b$, 1], and uses approximation of the function $\phi(x)$ with the corresponding polynomials in each of the regions. However, the author does not provide a way to handle the singularity problem of the logarithm function at the value $0$ in the interval [0, 1]. We propose to cater for this singularity problem by considering the notion of right-hand limit and continuity at one-sided open interval, and the notion of the improper integrals.
On the other hand, Orloff~\cite{orloff_uniqueness} provides the proof of the lemma by approximating the function $\phi(x)$ with a polynomial $p(x)$ in the interval [0, 1]. This can be achieved by either using the Stone-Weierstrass theorem~\cite{kantorovitz2003introduction} or by using the approximation of the function $\phi(x)$ with a polynomial $p(x)$ with respect to $L^2$ norm and is based on $L^p$ spaces. In this paper, we adopt the strategy based on $L^p$ spaces~\cite{arendt2011vector}  because of the availability of a rich formalization of $L^p$ spaces in HOL Light~\cite{hol_light2017lpspaces}. Whereas, in the case of Cohen's proof, we need to verify the properties of the improper integrals. The formal proof based on this strategy is more efficient, i.e., it requires less effort in the form of lines-of-code and man-hours as will be elaborated in Sections~\ref{SEC:lemma_lerch} and~\ref{SEC:Formalization_of_Lerchs_theorem}. Moreover, it is more generic than the other two methods, i.e., it considers an arbitrary interval [$a$, $b$] as the region of integration and thus can be directly used for the formal verification of the uniqueness property of the Fourier transform, which is our next goal.

The formalization presented in this paper is developed in higher-order logic (HOL) using the HOL Light theorem prover. The main motivation behind this choice is the availability of the multivariate calculus~\cite{harrison2013hol} (differentiation~\cite{hol_light2017diff}, integration~\cite{hol_light2017integ} and $L^p$ spaces~\cite{hol_light2017lpspaces}) and Laplace transform theories~\cite{taqdees2013formalization,rashid2017formal}. The proposed formalization is presented using a mix Math/HOL Light notation to make the paper easy to read for non-experts of HOL Light. The complete HOL Light script is available at~\cite{adnan16lerchsthm} for readers interested in viewing the HOL Light code.
In order to demonstrate the practical effectiveness of the Laplace transform theory in reasoning about the system analysis problems, we use it to conduct the formal analysis of a $4$-$\pi$ soft error crosstalk model, which is widely used in integrated circuits (ICs).

The rest of the paper is organized as follows: Section \ref{SEC:Preliminaries} provides a brief introduction about the HOL Light
theorem prover and the multivariable calculus theories of HOL Light, which act as preliminaries for the reported formalization. Section~\ref{SEC:form_laplace_transform} provides the formalization of the Laplace transform.
We describe the formalization of the lemma, given in Equations~(\ref{EQ:integral_lemma}) and~(\ref{EQ:conclusion_lemma}), for Lerch's theorem in Section~\ref{SEC:lemma_lerch}. Section~\ref{SEC:Formalization_of_Lerchs_theorem} presents the formalization of Lerch's theorem. Section~\ref{SEC:Application} presents our formal analysis of the soft error crosstalk model. Finally, Section~\ref{SEC:Conclusions} concludes the paper.


\section{Preliminaries} \label{SEC:Preliminaries}

In this section, we present an introduction to the HOL Light theorem prover and an overview about the multivariable calculus theories of HOL Light, which provide the foundational support for the proposed formalization.

\subsection{HOL Light Theorem Prover} \label{SUBSEC:HOL_Light_theorem_prover}

HOL Light~\cite{harrison-hol-light,harrison2009hol} is an interactive theorem proving environment for conducting proofs in higher-order logic.
The logic in the HOL Light system is represented in the strongly-typed functional programming language ML~\cite{paulson_96}.
Various mathematical foundations have been formalized and saved as HOL Light theories.
A HOL Light theory is a collection of valid HOL Light types, constants, axioms, definitions and theorems.
A theorem is a formalized statement that may be an axiom or could be deduced from already verified theorems by an inference rule.
It consists of a finite set $\Omega$ of Boolean terms, called the assumptions, and a Boolean term $S$, called the conclusion.
Soundness is assured as every new theorem must be verified by applying the basic axioms and primitive inference rules or any other previously verified theorems/inference rules.
The HOL Light theorem prover provides an extensive support of theorems regarding, boolean, arithmetics, real analysis and multivariate analysis in the form of theories, which are extensively used in our formalization. In fact, one of the primary reasons to chose the HOL Light theorem prover for the proposed formalization was the presence of an extensive support of multivariable calculus theories~\cite{hol_light2016multivariate}.

\subsection{Multivariable Calculus Theories in HOL Light} \label{SEC:Mult_cal_theories}

A $N$-dimensional vector is represented as a $\mathds{R}^N$ column matrix with each of its element as a real number in HOL Light~\cite{harrison2013hol,harrison2005hol}. All of the vector operations can thus be performed using matrix manipulations and all the multivariable calculus theorems are verified for functions with an arbitrary data-type $\mathds{R}^N \rightarrow \mathds{R}^M$. For example, a complex number is defined as a $2$-dimensional vector, i.e., a $\mathds{R}^2$ column matrix.

Some of the frequently used HOL Light functions in our work are explained below:


\begin{mdframed}
\begin{flushleft}
\begin{definition}
\label{DEF:cx_and_ii}
\emph{Cx and ii} \\{\small
\textup{\texttt{$\vdash$ $\forall$ a.\ Cx a = complex (a, \&0) \\
$\mathtt{}$$\vdash$ ii = complex (\&0, \&1)
}}}
\end{definition}
\end{flushleft}
\end{mdframed}

\noindent $\mathtt{Cx}$ is a type casting function from real ($\mathds{R}$) to complex ($\mathds{R}^2$),
whereas the $\texttt{\&}$ operator type casts a natural number ($\mathds{N}$) to its corresponding real number ($\mathds{R}$). Similarly, $\mathtt{ii}$ (iota) represents a complex number having the real part equal to zero and the magnitude of the imaginary part equal to $1$~\cite{harrison2007formalizing}.

\vspace*{0.2cm}

\begin{mdframed}
\begin{flushleft}
\begin{definition}
\label{DEF:re_im_lift_drop}
\emph{Re, Im, lift and drop} \\{\small
\textup{\texttt{$\vdash$ $\forall$ z.\ Re z = z\$1 \\
$\mathtt{}$$\vdash$ $\forall$ z.\ Im z = z\$2 \\
$\mathtt{}$$\vdash$ $\forall$ x.\ lift x = (lambda i.\ x) \\
$\mathtt{}$$\vdash$ $\forall$ x.\ drop x = x\$1
}}}
\end{definition}
\end{flushleft}
\end{mdframed}

\vspace*{0.2cm}

\noindent The functions $\mathtt{Re}$ and $\mathtt{Im}$ accept a complex number and return its real and imaginary part, respectively. Here, the notation $\mathtt{z\$i}$ represents the $i^{th}$ component of vector $\texttt{z}$. Similarly, the functions $\mathtt{lift}: \mathds{R} \rightarrow \mathds{R}^1$ and $\mathtt{drop}: \mathds{R}^1 \rightarrow \mathds{R}$ map a real number to a 1-dimensional vector and a $1$-dimensional vector to a real number, respectively~\cite{harrison2007formalizing}. Here, the function \texttt{lambda} is used to construct a vector componentwise~\cite{harrison2013hol}.

\vspace*{0.2cm}

\begin{mdframed}
\begin{flushleft}
\begin{definition}
\label{DEF:exp_ccos_csine}
\emph{Exponential, Complex Cosine and Sine} \\{\small
\textup{\texttt{$\vdash$ $\forall$ x.\ exp x = Re (cexp (Cx x)) \\
$\mathtt{}$$\vdash$ $\forall$ z.\ ccos z = (cexp (ii $\ast$ z) + cexp ($--$ii $\ast$ z)) / Cx (\&2)   \\
$\mathtt{}$$\vdash$ $\forall$ z.\ csin z = (cexp (ii $\ast$ z) - cexp ($--$ii $\ast$ z)) / (Cx (\&2) $\ast$ ii)
}}}
\end{definition}
\end{flushleft}
\end{mdframed}

\vspace*{0.2cm}

\noindent The complex exponential, real exponential, complex cosine and complex sine are represented as $\texttt{cexp}:\mathds{R}^2 \rightarrow \mathds{R}^2$, $\mathtt{exp}:\mathds{R} \rightarrow \mathds{R}$, $\mathtt{ccos}:\mathds{R}^2 \rightarrow \mathds{R}^2$ and $\mathtt{csin}:\mathds{R}^2 \rightarrow \mathds{R}^2$ in HOL Light, respectively~\cite{hol_light2016transcendentals}.

\vspace*{0.2cm}

\begin{mdframed}
\begin{flushleft}
\begin{definition}
\label{DEF:vector_integral}
\emph{Vector Integral and Real Integral} \\
{\small
\textup{\texttt{$\vdash$ $\forall$ f i.\ integral i f = (@y.\ (f has\_integral y) i)
}}} \\
{\small
\textup{\texttt{$\vdash$ $\forall$ f i.\ real\_integral i f = (@y.\ (f has\_real\_integral y) i) \\
}}}
\end{definition}
\end{flushleft}
\end{mdframed}

\vspace*{0.2cm}

\noindent The function $\mathtt{integral}$ represents the vector integral and is defined using the Hilbert choice operator $\texttt{@}$ in the functional form. It takes the integrand function $\texttt{f}: \mathds{R}^N \rightarrow \mathds{R}^M$, and a vector-space $\mathtt{i}: \mathds{R}^N \rightarrow \mathds{B}$, which defines the region of integration, and returns a vector $\mathds{R}^M$, which is the integral of $\mathtt{f}$ on $\mathtt{i}$. The function $\mathtt{has\_integral}$ represents the same relationship in the relational form. Similarly, the function $\mathtt{real\_integral}$ accepts an integrand function $\mathtt{f} : \mathds{R} \rightarrow \mathds{R}$ and a set of real numbers $\mathtt{i}: \mathds{R} \rightarrow \mathds{B}$ and returns the real-valued integral of the function $\mathtt{f}$ over $\mathtt{i}$.

\vspace*{0.2cm}

\begin{mdframed}
\begin{flushleft}
\begin{definition}
\label{DEF:vector_derivative}
\emph{Vector Derivative and Real Derivative} \\
{\small
\textup{\texttt{$\vdash$ $\forall$ f net.\ vector\_derivative f net =
}}} \\
{\small
\textup{\texttt{\hspace{4.0cm}(@f$'$.\ (f has\_vector\_derivative f$'$) net)
}}} \\
{\small
\textup{\texttt{$\vdash$ $\forall$ f x.\ real\_derivative f x =
}}} \\
{\small
\textup{\texttt{\hspace{4.0cm}(@f$'$.\ (f has\_real\_derivative f$'$) (atreal x))
}}}
\end{definition}
\end{flushleft}
\end{mdframed}

\vspace*{0.1cm}

\noindent The function $\mathtt{vector\_derivative}$ takes a function $\texttt{f} : \mathds{R}^1 \rightarrow \mathds{R}^M$ and a $\texttt{net} : \mathds{R}^1 \rightarrow \mathds{B}$, which defines the point at which $\texttt{f}$ has to be differentiated, and returns a vector of data-type $\mathds{R}^M$, which represents the differential of $\texttt{f}$ at $\texttt{net}$. The function $\mathtt{has\_vector\_derivative}$ defines the same relationship in the relational form.
Similarly, the function $\mathtt{real\_derivative}$ accepts a function $\texttt{f} : \mathds{R} \rightarrow \mathds{R}$ and a real number $\texttt{x}$, which represents the point where $\texttt{f}$ has to be differentiated, and returns the real-valued differential of $\texttt{f}$ at $\texttt{x}$.

\vspace*{0.2cm}

\begin{mdframed}
\begin{flushleft}
\begin{definition}
\label{DEF:limit_of_a vec_and_real_function}
\emph{Limit of a Vector and a Real function} \\{\small
\textup{\texttt{$\vdash$ $\forall$ f net.\ lim net f = (@l.\ (f $\rightarrow$ l) net)
}}} \\
{\small
\textup{\texttt{$\vdash$ $\forall$ f net.\ reallim net f = (@l.\ (f $\rightarrow$ l) net)
}}}
\end{definition}
\end{flushleft}
\end{mdframed}

\vspace*{0.1cm}

\noindent The function $\mathtt{lim}$ accepts a $\texttt{net}$ with elements of an arbitrary data-type $\mathds{A}$ and a function $\texttt{f} : \mathds{A} \rightarrow \mathds{R}^M$ and returns $\texttt{l}: \mathds{R}^M$, i.e., the value to which $\texttt{f}$ converges at the given $\texttt{net}$. Similarly, the function $\mathtt{reallim}$ accepts a $\texttt{net}$ with elements of data-type $\mathds{R}$ and a function $\texttt{f} : \mathds{R} \rightarrow \mathds{R}$ and returns $\texttt{l}: \mathds{R}$, i.e., the value to which $\texttt{f}$ converges at the given $\texttt{net}$.

In order to facilitate the understanding of the paper, we present the formalization of the Laplace transform, Lerch's theorem and the associated lemma using a mix Math/HOL Light notation. Some of the terms used, listed in Table~\ref{TAB:equivalent_math_conventions}, correlate with the traditional conventions, whereas the others are considered only to facilitate the understanding of this paper.

\vspace*{0.3cm}


\begin{footnotesize}
    \begin{longtable}{|p{1cm}|p{2cm}|p{2cm}|}
\caption{Conventions used for HOL Light Functions}
\label{TAB:equivalent_math_conventions}
\endfirsthead
\endhead
    \hline
    \hline
    \multicolumn{1}{l|}{\vtop{\hbox{\strut HOL Light}\hbox{\strut Functions}}}   &
    \multicolumn{1}{l|}{\vtop{\hbox{\strut Mathematical}\hbox{\strut Conventions}}}   &
    \multicolumn{1}{l}{\hspace{0.0cm} Description}

     \\ \hline \hline


\multicolumn{1}{l|}{  \texttt{lift x}   }  &
\multicolumn{1}{l|}{ $\mathtt{\overline{x}}$  }  &
\multicolumn{1}{l}{  \vtop{\hbox{\strut Conversion of a real number to $1$-dimensional}\hbox{\strut vector}}  }    \\ \hline

\multicolumn{1}{l|}{  \texttt{drop x}   }  &
\multicolumn{1}{l|}{ $\mathtt{\underline{x}}$  }  &
\multicolumn{1}{l}{  \vtop{\hbox{\strut Conversion of a $1$-dimensional vector to a real}\hbox{\strut number}}  }    \\ \hline

\multicolumn{1}{l|}{  \rule{0pt}{12pt} \hspace*{-0.2cm} \texttt{Cx a}   }  &
\multicolumn{1}{l|}{ $\mathtt{{\overrightarrow{\texttt{a}}}^2}$  }  &
\multicolumn{1}{l}{  Type casting from real ($\mathds{R}$) to complex ($\mathds{R}^2$) }    \\ \hline

\multicolumn{1}{l|}{  \texttt{exp x}   }  &
\multicolumn{1}{l|}{ $ \texttt{e}^{\texttt{x}} $ }  &
\multicolumn{1}{l}{  Real exponential function }    \\ \hline

\multicolumn{1}{l|}{ \rule{0pt}{12pt} \hspace*{-0.2cm}  \texttt{cexp x}   }  &
\multicolumn{1}{l|}{ $ \mathtt{{\overrightarrow{\texttt{e}}}}^{\texttt{x}} $  }  &
\multicolumn{1}{l}{  Complex exponential function }    \\ \hline

\multicolumn{1}{l|}{ \rule{0pt}{14pt} \hspace*{-0.2cm}  \texttt{integral}   }  &
\multicolumn{1}{l|}{ $\overrightarrow{\int}$  }  &
\multicolumn{1}{l}{  Integral of a vector-valued function }    \\ \hline

\multicolumn{1}{l|}{ \rule{0pt}{14pt} \hspace*{-0.2cm} \texttt{has\_integral}   }  &
\multicolumn{1}{l|}{ $\overrightarrow{\int}$  }  &
\multicolumn{1}{l}{  \vtop{\hbox{\strut Integral of a vector-valued function (Relational}\hbox{\strut  form)}}  }    \\ \hline

\multicolumn{1}{l|}{  \texttt{real\_integral}   }  &
\multicolumn{1}{l|}{ $\int$   }  &
\multicolumn{1}{l}{  Integral of a real-valued function }    \\ \hline

\multicolumn{1}{l|}{ \hspace*{-0.2cm}  \texttt{\vtop{\hbox{\strut has\_real\_integral}}}  }  &
\multicolumn{1}{l|}{ $\int$  }  &
\multicolumn{1}{l}{  \vtop{\hbox{\strut Integral of a real-valued function (Relational }\hbox{\strut  form)}}  }    \\ \hline

\multicolumn{1}{l|}{ \rule{0pt}{14pt} \hspace*{-0.2cm} $\texttt{lim}$   }  &
\multicolumn{1}{l|}{ $\mathbf{\overrightarrow{\lim}}$  }  &
\multicolumn{1}{l}{  Limit of a vector-valued function }    \\ \hline

\multicolumn{1}{l|}{ $\texttt{real\_lim}$   }  &
\multicolumn{1}{l|}{ $\lim$  }  &
\multicolumn{1}{l}{  Limit of a real-valued function }    \\ \hline

\multicolumn{1}{l|}{ $\mathtt{abs\ x}$   }  &
\multicolumn{1}{l|}{ $|\texttt{x}|$  }  &
\multicolumn{1}{l}{  Absolute value of a variable \texttt{x} }    \\ \hline

\multicolumn{1}{l|}{ \rule{0pt}{12pt} \hspace*{-0.2cm} $\mathtt{norm\ \overrightarrow{\mathtt{\texttt{x}}}}$ }  &
\multicolumn{1}{l|}{ $||{\mathtt{\overrightarrow{\texttt{x}}}}||$  }  &
\multicolumn{1}{l}{  Norm of a vector \texttt{x} }    \\ \hline

\multicolumn{1}{l|}{ \rule{0pt}{15pt} \hspace*{-0.2cm} \texttt{vsum}   }  &
\multicolumn{1}{l|}{ ${\overrightarrow{\mathtt{\sum}}}$  }  &
\multicolumn{1}{l}{  Summation of a vector-valued function }    \\ \hline

\multicolumn{1}{l|}{ \texttt{sum}   }  &
\multicolumn{1}{l|}{ $\mathtt{\sum}$  }  &
\multicolumn{1}{l}{  Summation of a real-valued function }    \\ \hline

\multicolumn{1}{l|}{ \hspace*{-0.2cm} \rule{0pt}{16pt} \hspace*{-0.2cm} \texttt{\vtop{\hbox{\strut vector\_derivati-}\hbox{\strut ve f (at t) }}}   }  &
\multicolumn{1}{l|}{ $\overrightarrow{\mathtt{\frac{df}{dt}}}$  }  &
\multicolumn{1}{l}{  Derivative of a vector-valued function \texttt{f} w.r.t \texttt{t} }    \\ \hline

\multicolumn{1}{l|}{ \hspace*{-0.2cm} \rule{0pt}{12pt} \hspace*{-0.2cm}  \texttt{\vtop{\hbox{\strut real\_derivative}\hbox{\strut f (at t) }}}  }  &
\multicolumn{1}{l|}{ $\mathtt{\frac{df}{dt}}$  }  &
\multicolumn{1}{l}{  Derivative of a real-valued function \texttt{f} w.r.t \texttt{t} }    \\ \hline

\multicolumn{1}{l|}{ \hspace*{-0.2cm} \rule{0pt}{16pt} \hspace*{-0.2cm}  \texttt{\vtop{\hbox{\strut higher\_vector\_de-}\hbox{\strut rivative n f t }}}   }  &
\multicolumn{1}{l|}{ $\overrightarrow{\mathtt{\frac{d^nf}{{dt}^n}}}$ }  &
\multicolumn{1}{l}{  \vtop{\hbox{\strut $n^{th}$ order derivative of a vector-valued function }\hbox{\strut  \texttt{f} w.r.t \texttt{t}}}    }    \\ \hline

\multicolumn{1}{l|}{ \hspace*{-0.2cm} \rule{0pt}{12pt} \hspace*{-0.2cm}  \texttt{\vtop{\hbox{\strut higher\_real\_deri-}\hbox{\strut vative n f t }}}    }  &
\multicolumn{1}{l|}{ $\mathtt{\frac{d^nf}{{dt}^n}}$  }  &
\multicolumn{1}{l}{  \vtop{\hbox{\strut $n^{th}$ order derivative of a real-valued function \texttt{f} }\hbox{\strut  w.r.t \texttt{t}}} }    \\ \hline


    \end{longtable}

\end{footnotesize}


\vspace*{0.2cm}

We build upon the above-mentioned fundamental functions of multivariable calculus in HOL Light to formalize the Laplace transform theory in the next sections.


\section{Formalization of the Laplace Transform} \label{SEC:form_laplace_transform}

Mathematically, the Laplace transform is defined for a function $f:\mathds{R}^1 \rightarrow \mathds{R}^2$ as~\cite{beerends2003fourier}:

\vspace*{0.2cm}

\begin{equation}\label{EQ:laplace_transform}
\mathcal{L} [f(t)] = F(s) = \int_{0}^{\infty} {f(t)e^{-s t}} dt, \ s \ \epsilon \  \mathds{C}
\end{equation}

\vspace*{0.2cm}

We formalize Equation~(\ref{EQ:laplace_transform}) in HOL Light as follows~\cite{rashid2017formal}:

\vspace*{0.2cm}

\begin{mdframed}
\begin{flushleft}
\begin{definition}
\label{DEF:laplace_transform_new}\emph{Laplace Transform} \\
{\small
\textup{\texttt{$\vdash$ $\forall$ s f.\ laplace\_transform f s = \\ \hspace*{5.0cm}  $\displaystyle \mathtt{\hspace{0.0cm} \overrightarrow{\int_{\mathtt{\gamma}}}{{\overrightarrow{\texttt{e}}}^{-s\mathtt{{\overrightarrow{\texttt{\big(\underline{t}\big)}}}^2}}}f(t)dt}$, $\mathtt{\gamma}$ = \{t | 0 $\leq$ \texttt{\underline{t}}\}
}}}
\end{definition}
\end{flushleft}
\end{mdframed}

\noindent The function \texttt{laplace\_transform} accepts a complex-valued function $ \texttt{f}: \mathds{R}^1 \rightarrow \mathds{R}^2 $
and a complex number $\texttt{s}$ and returns the Laplace transform of $ \texttt{f} $ as represented by
Equation~(\ref{EQ:laplace_transform}). In the above definition, we used the complex exponential function $ \mathtt{{\overrightarrow{\texttt{e}}}}: \mathds{R}^2
\rightarrow \mathds{R}^2 $ because the return data-type of the function $ \texttt{f} $ is $ \mathds{R}^2 $. Here, the data-type of $ \texttt{t} $ is $ \mathds{R}^1 $ and to multiply it with the complex number $ \mathtt{s} $, it is first converted into a real number \texttt{\underline{t}} by
using $\texttt{drop}$ and then it is converted to data-type $ \mathds{R}^2 $ using $ \texttt{Cx} $. Next, we use the vector
function $ \texttt{integral} $ (Definition~\ref{DEF:vector_integral}), i.e., $\overrightarrow{\int}$ to integrate the expression $ f(t)e^{-i \omega t} $ over
the positive real line since the data-type of this expression is $ \mathds{R}^2 $. The region of integration is
$\mathtt{\gamma}$, which represents the positive real line or the set \texttt{\{t | 0 $\leq$ \texttt{\underline{t}}\}}. The Laplace transform was earlier formalized using a limiting process as~\cite{taqdees2013formalization}:

\begin{mdframed}
\begin{flushleft}
{\small
\textup{\texttt{$\vdash$ $\forall$ s f.\ laplace\_transform f s =  $\displaystyle \mathtt{\hspace{0.0cm} \overrightarrow{\lim_{b \to \infty} } \overrightarrow{\int_{\mathtt{\overline{0}}}^{\mathtt{\overline{b}}}}{{e}^{-s\mathtt{{\overrightarrow{\texttt{\big(\underline{t}\big)}}}^2}}}f(t)dt}$
}}}
\end{flushleft}
\end{mdframed}


\noindent  However, the HOL Light definition of the integral function implicitly encompasses infinite limits of integration. So, our definition covers the region of integration, i.e., $[0, \infty)$, as
\texttt{\{t | 0 $\leq$ \texttt{\underline{t}}\}} and is equivalent to the definition given in~\cite{taqdees2013formalization}. However, our definition considerably simplifies the reasoning process in the verification of the Laplace transform properties since it does not involve the notion of limit.

The Laplace transform of a function $f$ exists, if $f$ is piecewise smooth and is of exponential order on the positive real
line~\cite{taqdees2013formalization,beerends2003fourier}.
A function is said to be piecewise smooth on an interval if it is piecewise
differentiable on that interval.

\begin{mdframed}
\begin{flushleft}
\begin{definition}
\label{DEF:laplace_existence}
\emph{Laplace Existence} \\{\small
\textup{\texttt{$\vdash$ $\forall$ s f.\ laplace\_exists f s = \\
$\mathtt{\ }$\hspace{0.6cm} $\displaystyle \mathtt{\Big(\forall\ b.\ f\ piecewise\_differentiable\_on\ \big[{\mathtt{\overline{0}}},\ {\mathtt{\overline{b}}}\big]\Big)\ \wedge}$ \\
$\mathtt{\ }$\hspace{0.6cm} $\mathtt{\Big(\exists\ M\ a.\ Re\big(s\big) > {\mathtt{\underline{a}}}\ \wedge\ exp\_order\_cond\ f\ M\ a\Big) }$
}}}
\end{definition}
\end{flushleft}
\end{mdframed}

\noindent The function $\texttt{exp\_order\_cond}$ in the above definition represents the exponential order condition necessary for the existence of the Laplace transform~\cite{taqdees2013formalization,beerends2003fourier}:

\begin{mdframed}
\begin{flushleft}
\begin{definition}
\label{DEF:exp_order_condition}
\emph{Exponential Order Condition} \\{\small
\textup{\texttt{$\vdash$ $\forall$ f M a. exp\_order\_cond f M a $\Leftrightarrow$  \\ \hspace*{2.5cm}
 $\displaystyle \mathtt{0 < M\ \wedge\ \Big(\forall\ t.\ 0 \le t \Rightarrow \big|\big|f \big(\mathtt{\overline{t}}\big)\big|\big| \le M\ e^{\mathtt{\underline{a}}t}\Big)} $
}}}
\end{definition}
\end{flushleft}
\end{mdframed}

We used Definitions~\ref{DEF:laplace_transform_new},~\ref{DEF:laplace_existence} and~\ref{DEF:exp_order_condition} to formally verify some of the classical properties of the Laplace transform, given in Table~\ref{TAB:properties of_lap_trans_new}. The properties namely linearity, frequency shifting, differentiation and integration were already verified using the formal definition of the Laplace transform~\cite{taqdees2013formalization}.
We formally verified these using our new definition of the Laplace transform~\cite{rashid2017formal}. Moreover, we formally verified some new properties, such as, time shifting, time scaling, cosine and sine-based modulations and the Laplace transform of a $n$-order differential equation~\cite{rashid2017formal}.
The assumptions of these theorems describe the existence of the corresponding Laplace transforms. For example, the predicate \texttt{laplace\_exists\_higher\_deriv} in the theorem corresponding to the $n$-order differential equation ensures that the Laplace of all the derivatives up to the $n^{th}$ order of the function \texttt{f} exist.  The function \texttt{diff\_eq\_n\_order} models the $n$-order differential equation itself.
Similarly, the predicate \texttt{differentiable\_higher\_derivative}  provides the differentiability of the function \texttt{f} and its higher derivatives up to the $n^{th}$ order.
Moreover, the HOL Light function \texttt{EL k lst} returns the $k^{th}$ element of a list \texttt{lst}.
The verification of these properties not only ensures the correctness of our definitions but also plays a vital role in minimizing the user effort in reasoning about the Laplace transform based analysis of systems, as will be depicted in Section~\ref{SEC:Application} of this paper.


\begin{footnotesize}
    \begin{longtable}{|p{2cm}|p{3cm}|p{7cm}|p{3cm}|}
\caption{Properties of the Laplace Transform}
\label{TAB:properties of_lap_trans_new}
\endfirsthead
\endhead
    \hline
    \hline
    \multicolumn{1}{l|}{Property}   &
    \multicolumn{1}{l}{\hspace{0.0cm} Formalized Form}

     \\ \hline \hline



\multicolumn{2}{c}{\textbf{Integrability}} \\ \hline

\multicolumn{1}{l|}{ {$\begin{array} {lcl}
    \hspace{0.0cm} \textit{$ e^{- s t} f(t)\ integrable\  $ } \\
\textit{$\mathtt{\ }$\hspace{0.4cm} $ on\ [0, \infty) $     }
 \end{array}$}  }  &

\multicolumn{1}{l}{{$\begin{array} {lcl} \textup{\texttt{\hspace{0.0cm}$\vdash$ $\forall$ f s.\ laplace\_exists f s     }} \\
\textup{\texttt{$\mathtt{\ }$\hspace{0.0cm} $\Rightarrow$ $\mathtt{{{\overrightarrow{e}}^{-s\mathtt{{\overrightarrow{\texttt{\big(\underline{t}\big)}}}^2}}}f(t)}$ integrable\_on \{t | 0 $\leq$ \underline{t}\}  }}
 \end{array}$}}    \\ \hline



\multicolumn{2}{c}{\textbf{Linearity}} \\ \hline

   \multicolumn{1}{l|}{ {$\begin{array} {lcl}
   \hspace{0.0cm} \textit{$ \mathcal{L} [ \alpha f(t) + \beta g(t)] = $ } \\
\textit{$\mathtt{\ }$\hspace{0.4cm} $\alpha F(s) + \beta G(s) $     }
 \end{array}$}  }  &

   \multicolumn{1}{l}{{$\begin{array} {lcl} \textup{\texttt{\hspace{0.0cm}$\vdash$ $\forall$ f g s a b.  }} \\
   \textup{\texttt{$\mathtt{\ }$ laplace\_exists f s $\wedge$ laplace\_exists g s  \hspace{-0.5cm}}} \\
\textup{\texttt{$\mathtt{\ }$\hspace{0.2cm} $\Rightarrow$  laplace\_transform \big(a $\ast$ $\mathtt{f(t)}$ + b $\ast$ $\mathtt{g(t)}$\big) s = \hspace{-0.5cm} }} \\
\textup{\texttt{$\mathtt{\ }$\hspace{1.20cm} a $\ast$ laplace\_transform f s +  \hspace{-0.5cm} }} \\
\textup{\texttt{$\mathtt{\ }$\hspace{2.00cm} b $\ast$ laplace\_transform g s \hspace{-0.5cm} }}
 \end{array}$}}    \\ \hline



\multicolumn{2}{c}{\textbf{Frequency Shifting}} \\ \hline

   \multicolumn{1}{l|}{ {$\begin{array} {lcl}
   \hspace{0.0cm} \textit{$ \mathcal{L} [ e^{s_0 t} f(t)] = $ } \\
\textit{$\mathtt{\ }$\hspace{0.4cm} $ F(s - s_0) $     }
 \end{array}$}  }  &

   \multicolumn{1}{l}{{$\begin{array} {lcl} \hspace{0.00cm} \textup{\texttt{$\vdash$ $\forall$ f s $\mathtt{s_0}$.\ laplace\_exists f s      \hspace{-0.5cm} }} \\
\textup{\texttt{$\mathtt{\ }$\hspace{-0.3cm} $\Rightarrow$ laplace\_transform \Big($\mathtt{{{\overrightarrow{e}}^{s_0\mathtt{{\overrightarrow{\texttt{\big(\underline{t}\big)}}}^2}}}f(t)}$\Big) s = \hspace{-0.5cm} }} \\
\textup{\texttt{$\mathtt{\ }$\hspace{2.4cm} laplace\_transform f (s - $\mathtt{s_0}$) \hspace{-0.5cm} }}
 \end{array}$}}    \\ \hline



\multicolumn{2}{c}{\textbf{First-order Differentiation in Time Domain}} \\ \hline

     \multicolumn{1}{l|}{{$\begin{array} {lcl}
  \hspace{0.0cm}  \mathcal{L} \left[\dfrac{d}{dt}f(t) \right] = \\
  \hspace{0.1cm}   s F(s) - f(0)
     \end{array}$}}    &

    \multicolumn{1}{l}{{$\begin{array} {lcl} \textup{\texttt{ \hspace{0.00cm} $\vdash$ $\forall$ f s.\ laplace\_exists f s $\wedge$  \hspace{-0.5cm} }} \\
\textup{\texttt{$\mathtt{\ }$\hspace{0.2cm} ($\forall$ t.\ f differentiable at t) $\wedge$  \hspace{-0.5cm} }}  \\
\textup{\texttt{$\mathtt{\ }$\hspace{0.0cm} laplace\_exists \bigg($\overrightarrow{\mathtt{\dfrac{df}{dt}}}$\bigg) s \hspace{-0.5cm}  }}  \\
\textup{\texttt{$\mathtt{\ }$\hspace{-0.2cm} $\Rightarrow$ laplace\_transform \bigg($\overrightarrow{\mathtt{\dfrac{df}{dt}}}$\bigg) s = \hspace{-0.5cm}  }}  \\
\textup{\texttt{$\mathtt{\ }$\hspace{0.8cm} s $\ast$ laplace\_transform f s - f \big($\mathtt{\overline{0}}$\big) \hspace{-0.5cm} }}
 \end{array}$}}    \\ \hline



\multicolumn{2}{c}{\textbf{Higher-order Differentiation in Time Domain}} \\ \hline

    \multicolumn{1}{l|}{{$\begin{array} {lcl}
  \hspace{0.0cm}  \mathcal{L} [\dfrac{d^n}{{dt}^n}f(t)] = s^n F(s) \\
  \hspace{-0.2cm}  - \sum_{k = 1}^{n}{ s^{k - 1} \dfrac{d^{n - k} f (0)}{{dx}^{n - k}} }
     \end{array}$}}    &

    \multicolumn{1}{l}{{$\begin{array} {lcl} \textup{\texttt{\hspace{0.00cm} $\vdash$ $\forall$ f s n. }} \\
\textup{\texttt{$\mathtt{\ }$\hspace{0.1cm} laplace\_exists\_higher\_deriv n f s $\wedge$ }} \\
\textup{\texttt{$\mathtt{\ }$\hspace*{-0.4cm} ($\forall$ t.\ differentiable\_higher\_derivative n f t) \hspace{-0.5cm}  }} \\
\textup{\texttt{$\mathtt{\ }$\hspace{0.1cm} $\Rightarrow$ laplace\_transform \bigg($\overrightarrow{\mathtt{\dfrac{d^nf}{{dt}^n}}}$\bigg) s = \hspace{-0.5cm} }} \\
\textup{\texttt{$\mathtt{\ }$\hspace{0.0cm} $\mathtt{s^n}$ $\ast$ laplace\_transform f s -  \hspace{-0.5cm} }} \\
\textup{\texttt{$\mathtt{\ }$\hspace{0.2cm} ${\overrightarrow{\mathtt{\sum_{k=1}^{n}}}}$
\bigg($\mathtt{s^{k-1}}$ $\overrightarrow{\mathtt{\dfrac{d^{n-k}f \big(\overline{0}\big)}{{dt}^{n-k}}}}$\bigg) }}
 \end{array}$}}    \\ \hline



\multicolumn{2}{c}{\textbf{Integration in Time Domain}} \\ \hline

    \multicolumn{1}{l|}{{$\begin{array} {lcl}
  \mathcal{L} \left[ \int_{0}^{t}{f (\tau) d\tau} \right] = \dfrac{1}{s} F(s)
     \end{array}$}}    &


    \multicolumn{1}{l}{{$\begin{array} {lcl} \textup{\texttt{\hspace{0.0cm}$\vdash$ $\forall$ f s.\ 0 < Re s $\wedge$ laplace\_exists f s $\wedge$   }}  \\
\textup{\texttt{$\mathtt{\ }$\hspace{0.0cm} laplace\_exists \bigg($\mathtt{\overrightarrow{\int_{\mathtt{\overline{0}}}^{\mathtt{t}}}\mathtt{f(\tau)d\tau}}$\bigg) s $\wedge$ \hspace{-0.5cm}  }}  \\
\textup{\texttt{$\mathtt{\ }$\hspace{0.0cm} \big($\forall$ x.\ f continuous\_on interval [$\mathtt{\overline{0}}$,x]\big) \hspace{-0.5cm}  }}  \\
\textup{\texttt{$\mathtt{\ }$\hspace{-0.2cm} $\Rightarrow$ laplace\_transform \bigg($\mathtt{\overrightarrow{\int_{\mathtt{\overline{0}}}^{\mathtt{t}}}\mathtt{f(\tau)d\tau}}$\bigg) s =  \hspace{-0.5cm} }}  \\
\textup{\texttt{$\mathtt{\ }$\hspace{2.0cm}  $\mathtt{\dfrac{\mathtt{{\overrightarrow{\texttt{1}}}^2}}{s}}$ $\ast$ laplace\_transform f s \hspace{-0.5cm}  }}
\vspace*{0.05cm}
\end{array}$}}    \\ \hline



\multicolumn{2}{c}{\textbf{Time Shifting}} \\ \hline

   \multicolumn{1}{l|}{ {$\begin{array} {lcl}
   \mathcal{L} \left[ f(t - t_0) u(t - t_0) \right] =  \\
   \textit{$\mathtt{\ }$\hspace{0.4cm} $  e^{-t_0 s} F(s) $    } \\

 \end{array}$}  }  &

    \multicolumn{1}{l}{{$\begin{array} {lcl} \textup{\texttt{\hspace{0.0cm} $\vdash$ $\forall$ f s $\mathtt{t_0}$.\ 0 < $\mathtt{\underline{t_0}}$ $\wedge$ laplace\_exists f s \hspace{-0.5cm}  }} \\
\textup{\texttt{$\mathtt{\ }$\hspace{0.1cm} $\Rightarrow$ laplace\_transform \big(shifted\_fun f $\mathtt{t_0}$\big) s = \hspace{-0.5cm}  }} \\
\textup{\texttt{$\mathtt{\ }$\hspace{0.6cm} $\mathtt{{\overrightarrow{e}}^{-s\mathtt{{\overrightarrow{\mathtt{\big(\underline{t_0}\big)}}}^2}}}$ $\ast$ laplace\_transform f s  \hspace{-0.5cm}  }}
 \end{array}$}}    \\ \hline



\multicolumn{2}{c}{\textbf{Time Scaling}} \\ \hline

    \multicolumn{1}{l|}{ {$\begin{array} {lcl}
     \mathcal{L} \left[ f(c t) \right] = \dfrac{1}{c} F\left(\dfrac{s}{c}\right),  \\
   \textit{$\mathtt{\ }$\hspace{0.4cm} $ \ \ 0 < c $  } \\
 \end{array}$}  }  &

    \multicolumn{1}{l}{{$\begin{array} {lcl} \textup{\texttt{$\vdash$   $\forall$ f s c.\ 0 < c $\wedge$ laplace\_exists f s $\wedge$    \hspace{-0.5cm}  }} \\
    \textup{\texttt{$\mathtt{\ }$\hspace{0.3cm} laplace\_exists f $\left(\mathtt{\dfrac{s}{\mathtt{{\overrightarrow{\texttt{c}}}^2}}}\right)$   \hspace{-0.5cm} }} \\
\textup{\texttt{$\mathtt{\ }$\hspace{0.0cm} $\Rightarrow$  laplace\_transform \big(f(c \% t)\big) s =  \hspace{-0.5cm} }} \\
\textup{\texttt{$\mathtt{\ }$\hspace{2.0cm} $\mathtt{\dfrac{\mathtt{{\overrightarrow{\texttt{1}}}^2}}{\mathtt{{\overrightarrow{\texttt{c}}}^2}}}$ $\ast$ laplace\_transform f $\left(\mathtt{\dfrac{s}{\mathtt{{\overrightarrow{\texttt{c}}}^2}}}\right)$  \hspace{-0.5cm} }}
\vspace*{0.05cm}
 \end{array}$}}    \\ \hline



\multicolumn{2}{c}{\textbf{Modulation (Cosine and Sine-based)}} \\ \hline

   \multicolumn{1}{l|}{ {$\begin{array} {lcl}
     \mathcal{L} \left[ f(t) cos(\omega_0t) \right] =  \\
   \textit{$\mathtt{\ }$\hspace{-0.1cm} $  \dfrac{F(s - i\omega_0)}{2} \ + $     } \\
   \textit{$\mathtt{\ }$\hspace{1.0cm} $ \dfrac{F(s + i\omega_0)}{2} $    }
 \end{array}$}  }  &

    \multicolumn{1}{l}{{$\begin{array} {lcl} \textup{\texttt{\hspace{0.0cm}$\vdash$   $\forall$ f s $\mathtt{w_0}$.\ laplace\_exists f s }} \\
\textup{\texttt{$\mathtt{\ }$\hspace{-0.25cm}  $\Rightarrow$  laplace\_transform     }}  \\
\textup{\texttt{$\mathtt{\ }$\hspace{0.5cm}   \Big(ccos \big($\mathtt{{\overrightarrow{\mathtt{w_0}}}^2}$$\ast$$\mathtt{{\overrightarrow{\texttt{\big(\underline{t}\big)}}}^2}$\big) f \big(t\big)\Big) s =  \hspace{-0.5cm}  }}  \\
\textup{\texttt{$\mathtt{\ }$\hspace{1.0cm}  $\mathtt{\dfrac{laplace\_transform\ f\ (s - ii \ast \mathtt{{\overrightarrow{\mathtt{w_0}}}^2})}{\mathtt{{\overrightarrow{\mathtt{2}}}^2}}\ +}$ \hspace{-0.4cm}  \hspace{-0.5cm} }} \\
\textup{\texttt{$\mathtt{\ }$\hspace{2.0cm}  $\mathtt{\dfrac{laplace\_transform\ f\ (s + ii \ast \mathtt{{\overrightarrow{\mathtt{w_0}}}^2})}{\mathtt{{\overrightarrow{\mathtt{2}}}^2}}}$ \hspace{-0.4cm}  \hspace{-0.5cm} }}
\vspace*{0.07cm}
 \end{array}$}}    \\ \hline


   \multicolumn{1}{l|}{ {$\begin{array} {lcl}
     \mathcal{L} \left[ f(t) sin(\omega_0t) \right] =  \\
   \textit{$\mathtt{\ }$\hspace{-0.1cm} $  \dfrac{F(s - i\omega_0)}{2i} \ - $     } \\
   \textit{$\mathtt{\ }$\hspace{1.0cm} $ \dfrac{F(s + i\omega_0)}{2i} $    }
 \end{array}$}  }  &

    \multicolumn{1}{l}{{$\begin{array} {lcl} \textup{\texttt{\hspace{0.0cm}$\vdash$   $\forall$ f s $\mathtt{w_0}$.\ laplace\_exists f s }} \\
\textup{\texttt{$\mathtt{\ }$\hspace{-0.25cm}  $\Rightarrow$  laplace\_transform     }}  \\
\textup{\texttt{$\mathtt{\ }$\hspace{0.5cm}   \Big(csin \big($\mathtt{{\overrightarrow{\mathtt{w_0}}}^2}$$\ast$$\mathtt{{\overrightarrow{\texttt{\big(\underline{t}\big)}}}^2}$\big) f \big(t\big)\Big) s =  \hspace{-0.5cm}  }}  \\
\textup{\texttt{$\mathtt{\ }$\hspace{1.0cm}  $\mathtt{\dfrac{laplace\_transform\ f\ (s - ii \ast \mathtt{{\overrightarrow{\mathtt{w_0}}}^2})}{\mathtt{{\overrightarrow{\mathtt{2}}}^2} \ast ii}\ -}$ \hspace{-0.4cm}  \hspace{-0.5cm} }} \\
\textup{\texttt{$\mathtt{\ }$\hspace{2.0cm}  $\mathtt{\dfrac{laplace\_transform\ f\ (s + ii \ast \mathtt{{\overrightarrow{\mathtt{w_0}}}^2})}{\mathtt{{\overrightarrow{\mathtt{2}}}^2} \ast ii}}$ \hspace{-0.4cm}  \hspace{-0.5cm} }}
\vspace*{0.07cm}
 \end{array}$}}    \\ \hline



\multicolumn{2}{c}{\textbf{$n$-order Differential Equation}} \\ \hline

   \multicolumn{1}{l|}{ {$\begin{array} {lcl}
     \mathcal{L} \Big( \sum _{k = 0}^{n} {{\alpha}_k \dfrac{d^ky}{{dt}^k}} \Big) =    \\
\textit{$\mathtt{\ }$\hspace{0.4cm} $F(s) \ \sum _{k = 0}^{n} {{\alpha}_k s^k} $    } \\
\textit{$\mathtt{\ }$\hspace{-0.2cm} $- \sum _{k = 0}^{n} {\sum_{i = 1}^{k}}$    } \\
\textit{$\mathtt{\ }$\hspace{0.5cm} ${s^{i - 1} \dfrac{d^{k - i}f(0)}{{dt}^{k - i}} }$    }
 \end{array}$}  }  &

    \multicolumn{1}{l}{{$\begin{array} {lcl} \textup{\texttt{$\vdash$ $\forall$ f lst s n.   }} \\
\textup{\texttt{$\mathtt{\ }$\hspace{0.1cm} laplace\_exists\_higher\_deriv n f s $\wedge$ }} \\
\textup{\texttt{$\mathtt{\ }$\hspace*{-0.4cm} ($\forall$ t.\ differentiable\_higher\_derivative n f t)  }} \\
\textup{\texttt{$\mathtt{\ }$\hspace{0.1cm} $\Rightarrow$ laplace\_transform    }} \\
\textup{\texttt{$\mathtt{\ }$\hspace{0.9cm} (diff\_eq\_n\_order n lst f t) s =   }} \\

\textup{\texttt{$\mathtt{\ }$\hspace*{-0.1cm} laplace\_transform f s $\ast$ ${\overrightarrow{\mathtt{\sum_{k=0}^{n}}}}$ \Big(EL k lst $\ast$ $\mathtt{s^k}$\Big) \hspace{-0.5cm}  }} \\

\textup{\texttt{$\mathtt{\ }$\hspace*{-0.3cm} - ${\overrightarrow{\mathtt{\sum_{k=0}^{n}}}}$ \Bigg(EL k lst $\ast$ ${\overrightarrow{\mathtt{\sum_{i=1}^{k}}}}$ \bigg($s^{i - 1}$ \bigg($\overrightarrow{\mathtt{\dfrac{d^{k-i}f \big(\overline{0}\big)}{{dt}^{k-i}}}}$\bigg)\bigg) \Bigg)  \hspace{-0.5cm} }}
\vspace*{0.07cm}
 \end{array}$}}    \\ \hline


    \end{longtable}

\end{footnotesize}


The generalized linear differential equation describes the input-output relationship for a generic $n$-order system~\cite{adams2012continuous}:

\small

\begin{equation}\label{EQ:diff_eqn_nth_order_LTI_sys}
 \sum _{k = 0}^{n} {{\alpha}_k \dfrac{d^k}{{dt}^k} y(t)} = \sum _{k = 0}^{m} {{\beta}_k \dfrac{d^k}{{dt}^k} x(t)}, \ \ \ \  m \leq n
\end{equation}

\normalsize

\noindent where $y(t)$ is the output and $x(t)$ is the input to the system. The constants $\alpha_k$ and
$\beta_k$ are the coefficients of the output and input differentials with order $k$, respectively. The greatest index $n$ of
the non-zero coefficient $\alpha_n$ determines the order of the underlying system.
The corresponding transfer function is obtained by setting the initial conditions equal to zero~\cite{nise2007control}:


\small

\begin{equation}\label{EQ:transfer_fun_nth_order_LTI_sys}
\dfrac{Y(s)}{X(s)} = \dfrac{\sum_{k = 0}^{m} {\beta_k s^k}}{\sum_{k = 0}^{n} {\alpha_k s^k}}
\end{equation}

\normalsize

We verified the transfer function, given in Equation~(\ref{EQ:transfer_fun_nth_order_LTI_sys}), for the generic $n$-order system as the following HOL Light theorem~\cite{rashid2017formal}.

\begin{mdframed}
	\begin{flushleft}
		\begin{theorem}
			\label{THM:transfer_fun_n_order_sys}
\emph{Transfer Function of a Generic $n$-order System} \\{\small
				\textup{\texttt{$\vdash$ $\forall$ y x m n inlst outlst s.   \\
						$\mathtt{\ }$\hspace{-0.1cm} ($\forall$ t.\ differentiable\_higher\_deriv m n x y t) $\wedge$  \\
						$\mathtt{\ }$\hspace{-0.1cm} laplace\_exists\_of\_higher\_deriv m n x y s $\wedge$ \\
						$\mathtt{\ }$\hspace{-0.1cm} zero\_init\_conditions m n x y $\wedge$  \\
						$\mathtt{\ }$\hspace{-0.1cm} diff\_eq\_n\_order\_sys m n inlst outlst y x $\wedge$  \\
						$\mathtt{\ }$\hspace{-0.1cm} laplace\_transform x s $\neq$ ${\mathtt{{\overrightarrow{\mathtt{0}}}^2}}$ $\wedge$ ${\overrightarrow{\mathtt{\sum_{k=0}^{n}}}}$ \Big(EL k outlst $\ast$ $\mathtt{s^k}$\Big) $\neq$ ${\mathtt{{\overrightarrow{\mathtt{0}}}^2}}$  \\
						$\mathtt{\ }$\hspace{1.0cm} $\Rightarrow$ $\mathtt{\dfrac{laplace\_transform\ y\ s}{laplace\_transform\ x\ s}}$ =       $\mathtt{\dfrac{{\overrightarrow{\mathtt{\sum_{k=0}^{m}}}}\ (EL\ k\ inlst\ \ast\ s^k)}{{\overrightarrow{\mathtt{\sum_{k=0}^{n}}}}\ (EL\ k\ outlst\ \ast\ s^k)}}$
			}}}
		\end{theorem}
	\end{flushleft}
\end{mdframed}

\noindent The first assumption ensures that the functions $\texttt{y}$ and $\texttt{x}$ are differentiable up to the $n^{th}$ and $m^{th}$
order, respectively. The next assumption represents the Laplace transform existence condition up to the $n^{th}$ order derivative
of function $\texttt{y}$ and $m^{th}$ order derivative of the function $\texttt{x}$. The next assumption models the zero initial
conditions for both of the functions $\texttt{y}$ and $\texttt{x}$, respectively. The next assumption represents the
formalization of Equation~(\ref{EQ:diff_eqn_nth_order_LTI_sys}) and the last two assumptions provide the conditions for the design of a reliable system.
Finally, the conclusion of the above theorem represents the transfer function given by Equation~(\ref{EQ:transfer_fun_nth_order_LTI_sys}).
The verification of this theorem is mainly based on \textit{$n$-order differential equation} property of the Laplace transform and is very useful as it allows to automate the verification of the transfer function of any system as will be seen in Section~\ref{SEC:Application} of the paper.
The formalization, described in this section, took around $2000$ lines of HOL Light code~\cite{adnan16lerchsthm} and around $110$ man-hours.


\section{Lemma for Lerch's Theorem} \label{SEC:lemma_lerch}

We formally verify the lemma (Equation~(\ref{EQ:conclusion_lemma})) involved in verifying Lerch's theorem for a function $f$ as the following HOL Light theorem:

\begin{mdframed}
\begin{flushleft}
\begin{theorem}
\label{THM:lerch_lemma_gen}
\emph{Lemma for a Vector-valued Function} \\{\small
\textup{\texttt{$\vdash$ $\forall$ f s. \\
$\mathtt{\ }$\hspace{0.5cm} bounded s $\wedge$   \\ \vspace*{0.15cm}
$\mathtt{\ }$\hspace{0.5cm} $\mathtt{\overline{{||f (x)||}^2}}$ integrable\_on s $\wedge$  \\ \vspace*{0.15cm}
$\mathtt{\ }$\hspace{0.5cm} \Big($\forall$ n.\ $\overrightarrow{\int_s}$$\mathtt{{\underline{x}}^n f(x)}$  = $\mathtt{{\overrightarrow{\texttt{0}}}^2}$\Big)  \\
$\mathtt{\ }$\hspace{2.5cm} $\Rightarrow$ negligible \big\{x | x IN s $\wedge$ f$($x$)$ $\neq$ $\mathtt{{\overrightarrow{\texttt{0}}}^2}$\big\}
}}}
\end{theorem}
\end{flushleft}
\end{mdframed}

The above theorem is the general version of the lemma (Equations~(\ref{EQ:integral_lemma}) and~(\ref{EQ:conclusion_lemma})) and it is verified for a vector-valued function \texttt{f} : $\mathds{R}^1 \rightarrow \mathds{R}^2$ and an arbitrary interval, i.e., set \texttt{s}. The first assumption of Theorem~\ref{THM:lerch_lemma_gen} ensures that the set \texttt{s} is bounded. The next assumption models the integrability condition for $\mathtt{\overline{{||f (x)||}^2}}$. The next assumption ensures that the integral of the complex integrand $x^nf(x)$ over the region of integration \texttt{s} is zero. Finally, the conclusion models the condition, which says that the size of the set containing all the values \texttt{x} $\mathtt{\in}$ \texttt{s} at which the function $f$ is zero is negligible. Alternatively, it means that the function $f$ is zero at every \texttt{x} $\mathtt{\in}$ \texttt{s}.
We proceed with the proof process of Theorem~\ref{THM:lerch_lemma_gen} by transforming the HOL Light function \texttt{negligible} into its counterpart for the real-valued functions, i.e., \texttt{real\_negligible}, which mainly requires the properties of vectors and negligible sets. Next, its proof is mainly based on the properties of integration along with the real-valued version of Theorem~\ref{THM:lerch_lemma_gen}, i.e., for the functions of data type $\mathds{R} \rightarrow \mathds{R}$, which is represented as:

\vspace*{0.2cm}

\begin{mdframed}
\begin{flushleft}
\begin{theorem}
\label{THM:lerch_lemma_gen_real}
\emph{Lemma for a Real-valued Function} \\{\small
\textup{\texttt{$\vdash$ $\forall$ f s.\   \\
$\mathtt{\ }$\hspace{0.5cm} real\_bounded s $\wedge$  \\
$\mathtt{\ }$\hspace{0.5cm} $\mathtt{{\big[f (x)\big]}^2}$ real\_integrable\_on s $\wedge$  \\
$\mathtt{\ }$\hspace{0.5cm} \Big($\forall$ n.\ $\int_s$$\mathtt{x^n f(x)}$  = $\mathtt{0}$\Big)  \\
$\mathtt{\ }$\hspace{2.5cm} $\Rightarrow$ real\_negligible \big\{x | x IN s $\wedge$ f$($x$)$ $\neq$ $\mathtt{0}$\big\}
}}}
\end{theorem}
\end{flushleft}
\end{mdframed}

\vspace*{0.2cm}

\noindent where all the assumptions of the above theorem are same as that of Theorem~\ref{THM:lerch_lemma_gen}. However, they hold for the real-valued function \texttt{f} : $\mathds{R} \rightarrow \mathds{R}$.
We start the proof process of the above theorem by converting the set \texttt{s} in to an interval, which directly implies from the first assumption of Theorem~\ref{THM:lerch_lemma_gen_real}, i.e., \texttt{real\_bounded s} and it results into the following subgoal:

\vspace*{0.2cm}

\begin{mdframed}
\begin{flushleft}
\begin{subgoal}
\label{SUBGOAL:lemma_real_neg_fun_not_eq_zero}
\emph{}
{\small\textup{\texttt{
$\mathtt{\ }$\hspace{0.0cm} real\_negligible \big\{x | x IN [a,b] $\wedge$ f$($x$)$ $\neq$ $\mathtt{0}$\big\}
}}}
\end{subgoal}
\end{flushleft}
\end{mdframed}

\vspace*{0.2cm}

Next, we assume $\mathtt{\overline{f(\underline{x})} = f'}$ and verify that the function $\mathtt{f'}$ belongs to the $L^2$ space, which is represented in HOL Light as:

\vspace*{0.2cm}

\begin{mdframed}
\emph{} 
{\small\textup{\texttt{
$\mathtt{\ }$\hspace{0.0cm} $\mathtt{f'}$ IN lspace ([$\overline{\texttt{a}}$,$\overline{\texttt{b}}$]) (2)
}}}
\end{mdframed}

\vspace*{0.2cm}

\noindent where the predicate \texttt{lspace} accepts a set (interval) $s$ and a real number $p$, which represents the order of the space and returns the corresponding $L^p$ space, i.e., it returns the set of functions $f$ such that each $f$ is measurable and $\overline{{||f(x)||}^2}$ is integrable on $s$. Its verification requires the properties of integration along with some real arithmetic reasoning and its serves as an assumption for the verification of Subgoal~\ref{SUBGOAL:lemma_real_neg_fun_not_eq_zero}.
Next, the following subgoal directly implies from Subgoal~\ref{SUBGOAL:lemma_real_neg_fun_not_eq_zero} as:

\vspace*{0.2cm}

\begin{mdframed}
\begin{flushleft}
\begin{subgoal}
\label{SUBGOAL:lemma_real_neg_fun_pow2_not_eq_zero}
\emph{} 
{\small\textup{\texttt{
$\mathtt{\ }$\hspace{0.0cm} real\_negligible \big\{x | x IN [a,b] $\wedge$ $\mathtt{[f(x)]^2}$ $\neq$ $\mathtt{{\overrightarrow{\texttt{0}}}^2}$\big\}
}}}
\end{subgoal}
\end{flushleft}
\end{mdframed}

\vspace*{0.2cm}

After applying the properties of the integrals and negligible sets along with some real arithmetic reasoning, it results into the following subgoal:

\vspace*{0.2cm}

\begin{mdframed}
\begin{flushleft}
\begin{subgoal}
\label{SUBGOAL:lemma_fun_pow2_le_e}
\emph{} 
{\small\textup{\texttt{
$\mathtt{\ }$\hspace{0.0cm} $\mathtt{\int_{a}^{b}}$$\mathtt{[f(x)]^2}$ $\le$ e
}}}
\end{subgoal}
\end{flushleft}
\end{mdframed}

\vspace*{0.2cm}

Now, the function \texttt{f} can be approximated by a polynomial \texttt{p$($x$)$} with respect to $L^2$ norm and we further verify:

\begin{mdframed}
{\small\textup{\texttt{
$\mathtt{\ }$\hspace{0.0cm} $\mathtt{\int_{a}^{b}}$$\mathtt{p(x) f(x)}$ = 0
}}}
\end{mdframed}

The above result after verification also serves as an assumption for Subgoal~\ref{SUBGOAL:lemma_fun_pow2_le_e}.
After applying transitivity property of real numbers, Subgoal~\ref{SUBGOAL:lemma_fun_pow2_le_e} results into the following subgoal:

\vspace*{0.2cm}

\begin{mdframed}
\begin{flushleft}
\begin{subgoal}
\label{SUBGOAL:lemma_fun)pow2_min_fun_mul_poly_eq_zero}
\emph{} 
{\small\textup{\texttt{
$\mathtt{\int_{a}^{b}}$$\mathtt{[f(x)]^2 dx}$ $\le$ $\mathtt{\int_{a}^{b}}$$\mathtt{\big([f(x)]^2 - p(x) f(x)\big) dx}$ $\wedge$ \\
\vspace*{0.2cm}
\hspace*{2.8cm} $\mathtt{\int_{a}^{b}}$$\mathtt{\big([f(x)]^2 - p(x) f(x)\big) dx}$ $\le$ e
}}}
\end{subgoal}
\end{flushleft}
\end{mdframed}

\vspace*{0.3cm}

The proof of the above subgoal is based on the properties of the integrals, $L^p$ spaces along with some real arithmetic reasoning. This concludes our proof of Theorem~\ref{THM:lerch_lemma_gen_real} and thus the lemma for Lerch's theorem. The details about the proof of the lemma can be found in the proof script~\cite{adnan16lerchsthm}.


\section{Formalization/ Formal Proof of Lerch's Theorem}\label{SEC:Formalization_of_Lerchs_theorem}

This section presents our formalization of Lerch's theorem using the HOL Light theorem prover. We formally verify the statement of Lerch's theorem as the following HOL Light theorem:

\vspace*{0.2cm}

\begin{mdframed}
\begin{flushleft}
\begin{theorem}
\label{THM:lerchs_theorem}
\emph{Lerch's Theorem} \\{\small
\textup{\texttt{$\vdash$ $\forall$ f g r.\  \\
$\mathtt{\ }$\hspace{0.5cm} 0 < \texttt{Re$($r$)$} $\wedge$  \\
$\mathtt{\ }$\hspace{0.5cm} \big($\forall$ s.\ \texttt{Re$($r$)$} $\le$ $\texttt{Re$($s$)$}$ $\Rightarrow$ laplace\_exists f s\big) $\wedge$ \\
$\mathtt{\ }$\hspace{0.5cm} \big($\forall$ s.\ $\texttt{Re$($r$)$}$ $\le$ $\texttt{Re$($s$)$}$ $\Rightarrow$ laplace\_exists g s\big) $\wedge$ \\
$\mathtt{\ }$\hspace{0.5cm} \big($\forall$ s.\ $\texttt{Re$($r$)$}$ $\le$ $\texttt{Re$($s$)$}$ $\Rightarrow$   \\
$\mathtt{\ }$\hspace{1.5cm} laplace\_transform f s = laplace\_transform g s\big)  \\
$\mathtt{\ }$\hspace{3.0cm} $\Rightarrow$ \Big($\forall$ t.\ 0 $\le$ $\mathtt{\underline{t}}$ $\Rightarrow$ f\big(t\big) = g\big(t\big)\Big)
}}}
\end{theorem}
\end{flushleft}
\end{mdframed}

\vspace*{0.2cm}

\noindent where \texttt{f} and \texttt{g} are vector-valued functions with data type $\mathds{R}^1 \rightarrow \mathds{R}^2$. Similarly, \texttt{r} and \texttt{s} are complex variables.
The first assumption of Theorem \ref{THM:lerchs_theorem} ensures the non-negativity of the real part of the Laplace variable $\texttt{r}$. The next two assumptions provide the Laplace existence conditions for the functions $\texttt{f}$ and $\texttt{g}$, respectively. The last assumption presents the condition that the Laplace transforms of the two complex-valued functions $\texttt{f}$ and $\texttt{g}$ are equal. Finally, the conclusion of Theorem \ref{THM:lerchs_theorem} presents the equivalence of the functions $\texttt{f}$ and $\texttt{g}$ for all values of their argument $t$ in $0 \le t$ since $t$ represents time that is always non-negative.
The proof of Theorem~\ref{THM:lerchs_theorem} mainly depends on the alternate representation of Lerch's theorem, which is verified as the following HOL Light theorem:

\vspace*{0.2cm}

\begin{mdframed}
\begin{flushleft}
\begin{theorem}
\label{THM:lerchs_theorem_alt}
\emph{Alternate Representation of Lerch's Theorem} \\{\small
\textup{\texttt{$\vdash$ $\forall$ f g N a b c. \\
$\mathtt{\ }$\hspace{0.5cm} a + 1 < N $\wedge$  \\ \vspace*{0.1cm}
$\mathtt{\ }$\hspace{0.5cm} f continuous\_on \big\{t | 0 $\leq$ $\mathtt{\underline{t}}$\big\} $\wedge$  \\ \vspace*{0.1cm}
$\mathtt{\ }$\hspace{0.5cm} g continuous\_on \big\{t | 0 $\leq$ $\mathtt{\underline{t}}$\big\} $\wedge$  \\ \vspace*{0.1cm}
$\mathtt{\ }$\hspace{0.5cm} \Big($\forall$ t.\ 0 $\leq$ $\mathtt{\underline{t}}$  $\Rightarrow$ \big|\big|f$($t$)$\big|\big| $\leq$ b$\mathtt{e^{a\underline{t}}}$  $\wedge$  \big|\big|g$($t$)$\big|\big| $\leq$ c$\mathtt{e^{a\underline{t}}}$\Big)  $\wedge$   \\ \vspace*{0.1cm}
$\mathtt{\ }$\hspace{0.5cm} \big($\forall$ n.\ N $\le$ n $\Rightarrow$   \\
$\mathtt{\ }$\hspace{1.5cm} laplace\_transform f $\mathtt{{\overrightarrow{\texttt{n}}}^2}$ = laplace\_transform g $\mathtt{{\overrightarrow{\texttt{n}}}^2}$\big)  \\
$\mathtt{\ }$\hspace{3.0cm} $\Rightarrow$ \Big($\forall$ t.\ 0 $\le$ $\mathtt{\underline{t}}$ $\Rightarrow$ f\big(t\big) = g\big(t\big)\Big)
}}}
\end{theorem}
\end{flushleft}
\end{mdframed}

\vspace*{0.2cm}


\noindent
where the first assumption models the upper bound of the exponent \texttt{a} of the exponential function. The next two assumptions provide the continuity of the complex-valued functions \texttt{f} and \texttt{g} over the interval $[0, \infty)$, respectively. The next assumption presents the upper bounds of the functions \texttt{f} and \texttt{g}, which is very similar to the exponential order condition (Definition~\ref{DEF:exp_order_condition}).
The last assumption describes the condition that the Laplace transforms of the two functions \texttt{f} and \texttt{g} are equal. Finally, the conclusion presents the equivalence of the functions \texttt{f} and \texttt{g}.
We proceed with the proof of Theorem~\ref{THM:lerchs_theorem_alt} by applying the properties of sets along with some complex arithmetic simplification, which results into the following subgoal:

\vspace*{0.2cm}

\begin{mdframed}
\begin{flushleft}
\begin{subgoal}
\label{SUBGOAL:thm_alt_rep_lerch_set_conv}
\emph{} 
{\small\textup{\texttt{
$\mathtt{\ }$\hspace{0.0cm} $\forall$ t.\ t IN \big\{x | 0 $\le$ $\underline{\texttt{x}}$\big\} $\Rightarrow$ f$($t$)$ - g$($t$)$ = $\mathtt{{\overrightarrow{\texttt{0}}}^2}$
}}}
\end{subgoal}
\end{flushleft}
\end{mdframed}

\vspace*{0.2cm}

The proof of the above subgoal is mainly based on the following lemma:

\vspace*{0.2cm}

\begin{mdframed}
\begin{flushleft}
\begin{lemma}
\label{LEM:neglig_conti_preimage_lemma}
{\small
\textup{\texttt{$\vdash$ $\forall$ f s a.\ convex s $\wedge$  \\ \vspace*{0.1cm}
$\mathtt{\ }$\hspace{1.0cm} \big(interior s = \{\} $\Rightarrow$ s = \{\}\big) $\wedge$  \\ \vspace*{0.1cm}
$\mathtt{\ }$\hspace{1.0cm} f continuous\_on s $\wedge$  \\ \vspace*{0.1cm}
$\mathtt{\ }$\hspace{1.0cm} negligible \big\{x | x IN s $\wedge$ f$($x$)$ $\neq$ a\big\}  \\ \vspace*{0.1cm}
$\mathtt{\ }$\hspace{3.5cm} $\Rightarrow$  \big($\forall$ x.\ x IN s $\Rightarrow$ f$($x$)$ = a\big)
}}}
\end{lemma}
\end{flushleft}
\end{mdframed}

\vspace*{0.2cm}

The application of the above lemma on Subgoal~\ref{SUBGOAL:thm_alt_rep_lerch_set_conv} results into a subgoal, where it is required to verify all the assumptions of Lemma~\ref{LEM:neglig_conti_preimage_lemma}. The first three assumptions are verified using the properties of continuity and sets along with some complex arithmetic reasoning. Finally, the fourth assumption results into the following subgoal:

\vspace*{0.2cm}

\begin{mdframed}
\begin{flushleft}
\begin{subgoal}
\label{SUBGOAL:neg_pos_line_fun_eq_zero}
\emph{} 
{\small\textup{\texttt{
$\mathtt{\ }$\hspace{0.0cm} negligible \Big\{t | 0 $\le$ $\underline{\texttt{t}}$ $\wedge$ \big(f$($t$)$ - g$($t$)$\big) $\ne$ $\mathtt{{\overrightarrow{\texttt{0}}}^2}$\Big\}
}}}
\end{subgoal}
\end{flushleft}
\end{mdframed}

\vspace*{0.2cm}

The proof of the above subgoal is mainly based on the following theorem by setting the value of the function $h(t) = f(t) - g(t)$:

\vspace*{0.2cm}

\begin{mdframed}
\begin{flushleft}
\begin{theorem}
\label{THM:lerch_zero}
\emph{Generalization of Lerch's Theorem} \\{\small
\textup{\texttt{$\vdash$ $\forall$ h s a.\   \\
$\mathtt{\ }$\hspace{0.0cm} h measurable\_on \big\{t | 0 $\leq$ $\mathtt{\underline{t}}$\big\} $\wedge$   \\ \vspace*{0.05cm}
$\mathtt{\ }$\hspace{0.0cm} a + 1 < N $\wedge$    \\ \vspace*{0.05cm}
$\mathtt{\ }$\hspace{0.0cm} \big($\forall$ t.\ 0 $\leq$ $\mathtt{\underline{t}}$  $\Rightarrow$ \big|\big|h$($t$)$\big|\big| $\leq$ b$\mathtt{e^{a\underline{t}}}$\big) $\wedge$  \\ \vspace*{0.1cm}
$\mathtt{\ }$\hspace{0.0cm} \big($\forall$ n.\ N $\le$ n $\Rightarrow$ laplace\_transform h $\mathtt{{\overrightarrow{\texttt{n}}}^2}$ =  $\mathtt{{\overrightarrow{\texttt{0}}}^2}$\big)  \\
$\mathtt{\ }$\hspace{2.5cm} $\Rightarrow$ negligible \big\{t | 0 $\leq$ $\mathtt{\underline{t}}$ $\wedge$ h$($t$)$ $\neq$ $\mathtt{{\overrightarrow{\texttt{0}}}^2}$\big\}
}}}
\end{theorem}
\end{flushleft}
\end{mdframed}


\noindent where the first assumption models the condition that the function \texttt{h} is measurable on the interval $[0, \infty)$. The next two assumptions provide the upper bounds of the exponent \texttt{a} and the complex-valued function \texttt{h}.
The last assumption describes the condition that the Laplace transform of the function \texttt{h} is equal to zero. Finally, the conclusion uses the predicate \texttt{negligible} to model the condition that the function \texttt{h(t)} is equal to zero.
We proceed with the proof of Theorem~\ref{THM:lerch_zero} by verifying the following subgoal:

\vspace*{0.2cm}

\begin{mdframed}
\begin{flushleft}
\begin{subgoal}
\label{SUBGOAL:gn_meas_norm_gn_ubounded}
\emph{} 
{\small\textup{\texttt{
$\mathtt{\ }$\hspace{0.0cm} \Big($\forall$ n.\ N $\le$ n $\Rightarrow$ g n measurable\_on \big($\overline{\texttt{0}}$,$\overline{\texttt{1}}$\big)\Big) $\wedge$ \\ \vspace*{0.1cm}
$\mathtt{\ }$\hspace{0.0cm} \Big($\forall$ n x.\ N $\le$ n $\wedge$ x IN \big($\overline{\texttt{0}}$,$\overline{\texttt{1}}$\big) $\Rightarrow$ \big|\big|g n x\big|\big| $\le$ b\Big)
}}}
\end{subgoal}
\end{flushleft}
\end{mdframed}

\vspace*{0.2cm}

\noindent where,

\vspace*{0.2cm}

{\small\textup{\texttt{
$\mathtt{\ }$\hspace{0.0cm} g = h\bigg(--$\mathtt{\overline{(log (\underline{x}))}}$\bigg) $\bigg(\mathtt{{\overrightarrow{\texttt{\big(\underline{x}\big)}}}^2}\bigg)^{\mathtt{\mathtt{{\overrightarrow{\texttt{n - 1}}}^2}}}$
}}}

\vspace*{0.2cm}

The proof of the above subgoal is mainly based on applying cases on \texttt{N $\le$ n} along with the following lemma:

\vspace*{0.2cm}

\begin{mdframed}
\begin{flushleft}
\begin{lemma}
\label{LEM:laplace_trans_bounded_measurable}
{\small
\textup{\texttt{$\vdash$ $\forall$ h a b s.  \\ \vspace*{0.05cm}
$\mathtt{\ }$\hspace{0.5cm} h measurable\_on \big\{t | 0 $\leq$ $\mathtt{\underline{t}}$\big\} $\wedge$  \\ \vspace*{0.05cm}
$\mathtt{\ }$\hspace{0.5cm} a + 1 < Re$($s$)$ $\wedge$  \\  \vspace*{0.1cm}
$\mathtt{\ }$\hspace{0.5cm} \big($\forall$ t.\ 0 $\leq$ $\mathtt{\underline{t}}$  $\Rightarrow$ \big|\big|h$($t$)$\big|\big| $\leq$ b$\mathtt{e^{a\underline{t}}}$\big)   \\ \vspace*{0.1cm}
$\mathtt{\ }$\hspace{0.5cm} $\Rightarrow$  h\bigg(--$\mathtt{\overline{(log (\underline{x}))}}$\bigg)$\bigg(\mathtt{{\overrightarrow{\texttt{\big(\underline{x}\big)}}}^2}\bigg)^{\mathtt{s - \mathtt{{\overrightarrow{\texttt{1}}}^2}}}$ measurable\_on \big($\overline{\texttt{0}}$,$\overline{\texttt{1}}$\big) $\wedge$  \\ \vspace*{0.05cm}
$\mathtt{\ }$\hspace{1.0cm} \Bigg($\forall$ x. x IN \big($\overline{\texttt{0}}$,$\overline{\texttt{1}}$\big) $\Rightarrow$  \bigg|\bigg| h\bigg(--$\mathtt{\overline{(log (\underline{x}))}}$\bigg)$\bigg(\mathtt{{\overrightarrow{\texttt{\big(\underline{x}\big)}}}^2}\bigg)^{\mathtt{s - \mathtt{{\overrightarrow{\texttt{1}}}^2}}}$ \bigg|\bigg| $\leq$ b \Bigg)
}}}
\end{lemma}
\end{flushleft}
\end{mdframed}

\vspace*{0.2cm}

The singularity of the logarithm function at value $0$ in the above lemma is handled by taking the measurability of the function $h (-log\ x) x^{(s-1)}$ over the interval (0, 1).
The verification of Subgoal~\ref{SUBGOAL:gn_meas_norm_gn_ubounded} serves as one of the assumption for the verification of Theorem~\ref{THM:lerch_zero}. Next, we simplify the conclusion of Theorem~\ref{THM:lerch_zero} using all the assumptions and properties of the sets, to obtain the following subgoal:

\vspace*{0.2cm}

\begin{mdframed}
\begin{flushleft}
\begin{subgoal}
\label{SUBGOAL:subgoal_for_main_lemma_appli}
\emph{} 
{\small\textup{\texttt{
$\mathtt{\ }$\hspace{0.0cm} negligible \big\{x | x IN ($\overline{\texttt{0}}$ $\overline{\texttt{1}}$) $\wedge$ g N x $\neq$ $\mathtt{{\overrightarrow{\texttt{0}}}^2}$\big\}
}}}
\end{subgoal}
\end{flushleft}
\end{mdframed}

\vspace*{0.2cm}

The proof of the above subgoal is mainly based on the main lemma (Theorem~\ref{THM:lerch_lemma_gen}), properties of integration and sets along with some complex arithmetic reasoning. This concludes our formal proof of Lerch's theorem.

Our proof script of the formalization, presented in Sections \ref{SEC:lemma_lerch} and \ref{SEC:Formalization_of_Lerchs_theorem}, consists of about 700 lines-of-code and it took about 45 man-hours for the verification.
One of the major difficulties faced in the reported formalization was
the unavailability of a formal proof for Lerch's theorem. Most of the mathematical texts on Laplace transform, e.g.,~\cite{beerends2003fourier} and~\cite{schiff2013laplace}, mention the uniqueness property of the Laplace transform without presenting its proof. We only found a couple of analytical paper-and-pencil proofs~\cite{cohen2007numerical,orloff_uniqueness} of Lerch's theorem, which formed the basis of the reported formalization.
Secondly, we verified Lerch's theorem for the complex-valued function ($\mathcal{L} [f (t)]$ or $F(s)$), whereas the available paper-and-pencil proofs~\cite{cohen2007numerical,orloff_uniqueness} were based on a real-valued function.
The formalization of Lerch's theorem enabled us to formally verify the solutions of the differential equations, which was not possible using the formalization of the Laplace transform presented in~\cite{taqdees2013formalization,rashid2017formal}. We illustrate the practical effectiveness of our formalized Laplace transform theory by presenting the formal analysis of a $4$-$\pi$ soft error crosstalk model for ICs in the following section.


\section{Formal Analysis of a $4$-$\pi$ Soft Error Crosstalk Model for Nanometer Technologies}\label{SEC:Application}

With the advancement in the Complementary Metal-oxide Semiconductor (CMOS) technologies, nanometer circuits are becoming more vulnerable to soft errors, such as, clock jitters~\cite{seifert2005radiation}, soft delays~\cite{gill2004soft}, coupling noise, crosstalk noise pulses that are caused by Single Event (SE) particles~\cite{sayil2016soft}, signal cross-coupling effects~\cite{balasubramanian2006crosstalk,sayil2013single} and voltage drops in power supply, and can badly effect the integrity of the signals. These circuits usually contain a huge amount of interconnection lines, in addition to the transistors, due to the scaling down of the deep submicron CMOS technology. Moreover, these lines can interfere with each other, contributing to the degradation of the performance of the circuit and thus cannot be considered as electrically isolated components. The increase in the heights of wires and reduction in the distances between the adjacent wires are the main causes of this interference, which can result in to crosstalk noise and signal delays.
Modeling of these crosstalk noise and delays caused by SE particles and other sources can be helpful in identifying them and also in rectifying their effects on the CMOS technology. It also enables the designers to develop a low-power and energy efficient CMOS circuit. Due to the wider utility of CMOS technologies in safety and mission critical applications, such as medicine~\cite{bradley1998single}, military~\cite{schrimpf2004radiation} and avionics~\cite{normand1996single}, the formal modeling and analysis of the soft error crosstalk in these technologies is of utmost importance as the verification of these models enhances the reliability and security of the overall system.

\begin{figure}[ht!]
\centering
\scalebox{0.55}
{\includegraphics[trim={2.0 0.5cm 2.0 0.5cm},clip]{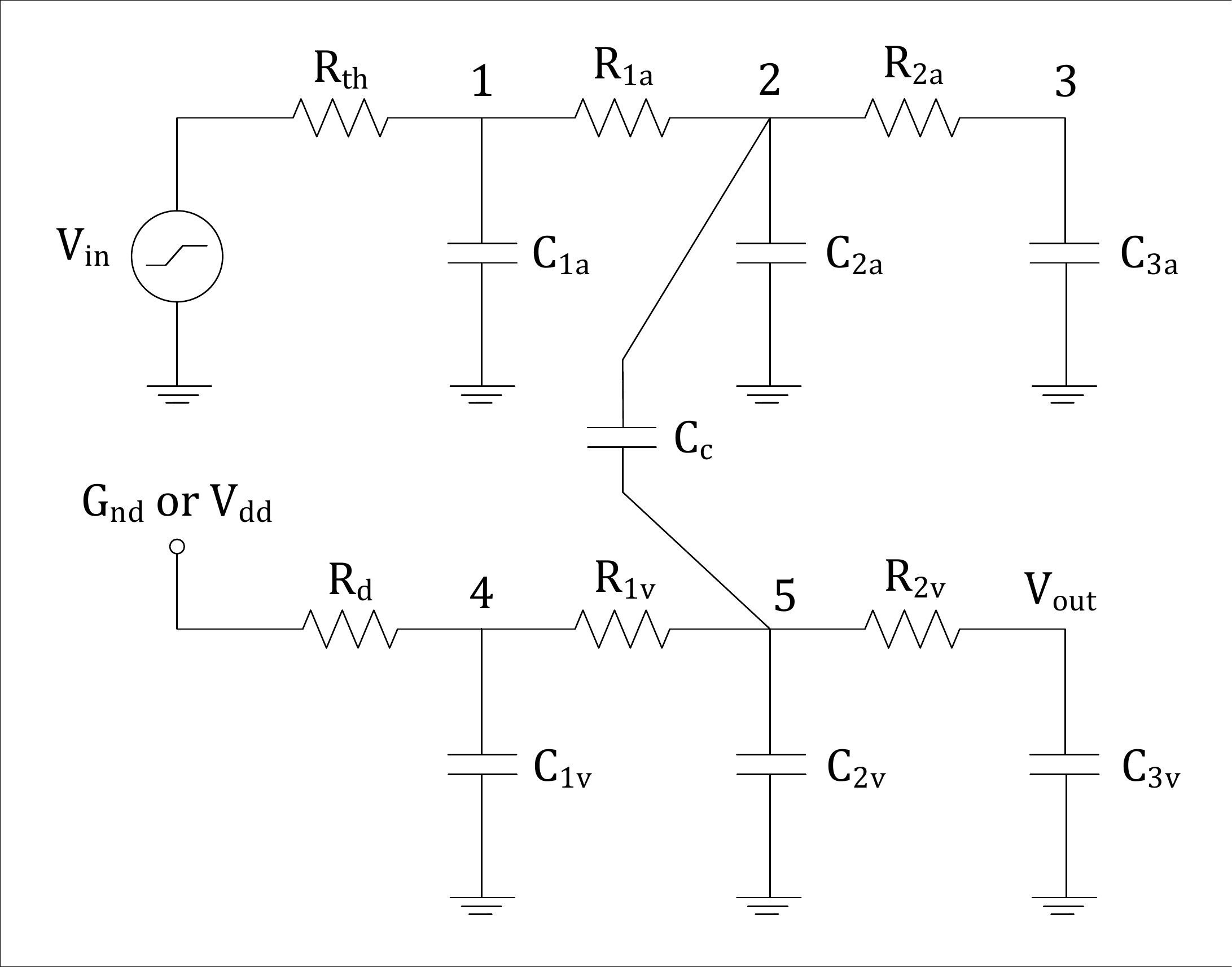}}
\caption{$4$-$\pi$ Interconnect Circuit Modeling the SE Crosstalk Effect~\cite{sayil2016soft}}
\label{FIG:4pimodel}
\end{figure}

A $4$-$\pi$ interconnect circuit, depicted in Figure~\ref{FIG:4pimodel}, models the SE crosstalk effect in the CMOS technologies~\cite{sayil2016soft,sayil2007precise}. It mainly consists of two $2$-$\pi$ circuits that model the aggressor and victim lines (nets), respectively. Here, $R_{1a}$ and $R_{2a}$ are the resistors corresponding to the aggressor net, whereas, $C_{1a}$, $C_{2a}$ and $C_{3a}$ are the respective capacitors. Similarly, in the case of the victim net, $R_{1v}$ and $R_{2v}$ are the resistors, and $C_{1v}$, $C_{2v}$ and $C_{3v}$ are the respective capacitors. Also, $C_c$ is the coupling capacitor used between the aggressor and the victim nets.

\subsection{Formal Analysis of Passive Aggressor}\label{SUBSEC:form_anal_pass_aggress}

Based on the $4$-$\pi$ interconnect circuit (Figure~\ref{FIG:4pimodel}), the passive aggressive model for analyzing the crosstalk noise and delay, is depicted in Figure~\ref{FIG:aggressor_model}, which is obtained as a result of applying the decoupling approach~\cite{sayil2016soft,sayil2007precise}. The resistance $R_{th}$ is the effective resistance of the aggressor driver~\cite{sayil2016soft}.
For the analysis of the passive aggressor, we first need to formalize its dynamical behaviour in the form of its governing differential equation in higher-order logic.
We use the generic differential equation of order $n$, to model the differential equation of the passive aggressor as follows:

\begin{figure}[H]
\centering
\scalebox{0.55}
{\includegraphics[trim={2.0 0.5cm 2.0 0.5cm},clip]{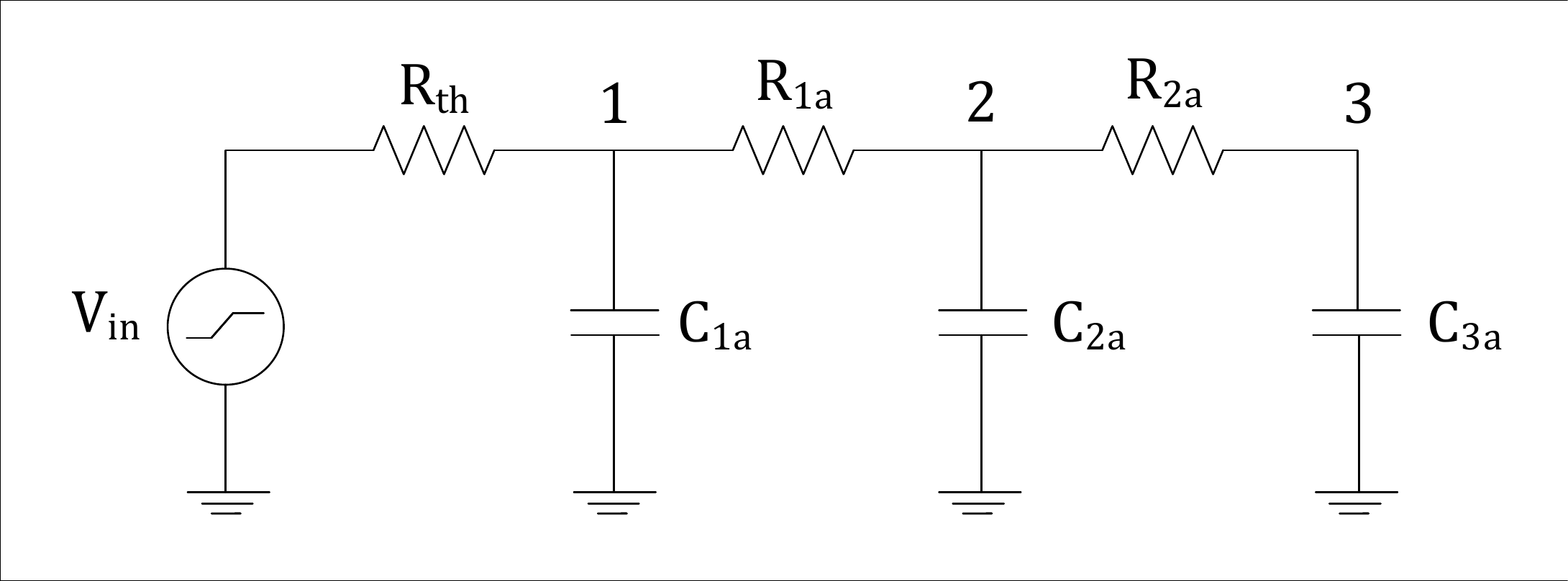}}
\caption{Passive Aggressor Model~\cite{sayil2007precise}}
\label{FIG:aggressor_model}
\end{figure}

\vspace*{0.1cm}


\begin{mdframed}
\begin{flushleft}
\begin{definition}
\label{DEF:pass_aggre_behav_model}
\emph{Behavioural Specification of Passive Aggressor} \\{
\small
\textup{\texttt{$\vdash$ $\forall$ $\mathtt{R_{1a}\ R_{2a}\ C_{2a}\ C_{3a}}$.   }}} \\ \vspace*{0.04cm}
$\mathtt{\ }$\hspace{0.8cm}  \small\textup{\texttt{inlst\_pass\_aggres $\mathtt{R_{1a}\ R_{2a}\ C_{2a}\ C_{3a}}$ = \big[$\mathtt{{\overrightarrow{\texttt{1}}}^2}$; $\mathtt{{\overrightarrow{\texttt{A}}}^2}$; $\mathtt{{\overrightarrow{\texttt{B}}}^2}$; $\mathtt{{\overrightarrow{\texttt{C}}}^2}$\big]}} \\ \vspace*{0.04cm}
{\small
\textup{\texttt{$\vdash$ $\forall$ $\mathtt{R_{1a}\ R_{2a}\ R_{th}\ C_{1a}\ C_{2a}\ C_{3a}}$.  }}}\\ \vspace*{0.04cm}
$\mathtt{\ }$\hspace{0.8cm}  \small\textup{\texttt{outlst\_pass\_aggres $\mathtt{R_{1a}\ R_{2a}\ R_{th}\ C_{1a}\ C_{2a}\ C_{3a}}$ = }} \\ \vspace*{0.04cm}
$\mathtt{\ }$\hspace{5.0cm}  \small\textup{\texttt{ \big[$\mathtt{{\overrightarrow{\texttt{1}}}^2}$; $\mathtt{{\overrightarrow{\texttt{D}}}^2}$; $\mathtt{{\overrightarrow{\texttt{E}}}^2}$; $\mathtt{{\overrightarrow{\texttt{F}}}^2}$; $\mathtt{{\overrightarrow{\texttt{G}}}^2}$; $\mathtt{{\overrightarrow{\texttt{H}}}^2}$\big]}} \\ \vspace*{0.04cm}
{\small
\textup{\texttt{$\vdash$ $\forall$ $\mathtt{R_{th}\ C_{1a}\ V_2\ R_{1a}\ R_{2a}\ C_{2a}\ C_{3a}\ V_{in}}$ t. \\ \vspace*{0.04cm} \vspace*{0.04cm}
$\mathtt{\ }$\hspace{0.8cm}  pass\_aggressor\_behav\_spec $\mathtt{R_{1a}\ R_{2a}\ R_{th}\ C_{1a}\ C_{2a}\ C_{3a}\ V_{in}\ V_2}$ t $\Leftrightarrow$ \\ \vspace*{0.04cm}
$\mathtt{\ }$\hspace{0.8cm}  diff\_eq\_n\_order 5  \\ \vspace*{0.04cm}
$\mathtt{\ }$\hspace{2.0cm}  (outlst\_pass\_aggres $\mathtt{R_{1a}\ R_{2a}\ R_{th}\ C_{1a}\ C_{2a}\ C_{3a}}$) $\mathtt{V_2}$ t = \\ \vspace*{0.04cm}
$\mathtt{\ }$\hspace{0.8cm}  diff\_eq\_n\_order 3   \\ \vspace*{0.04cm}
$\mathtt{\ }$\hspace{2.0cm}  (inlst\_pass\_aggres $\mathtt{R_{1a}\ R_{2a}\ C_{2a}\ C_{3a}}$) $\mathtt{V_{in}}$ t
}}}
\end{definition}
\end{flushleft}
\end{mdframed}

\vspace*{0.2cm}

\noindent where $\mathtt{V_{in}}$ is the input voltage having data type $\mathtt{\mathds{R}^1 \rightarrow \mathds{R}^2}$. Similarly, $\mathtt{V_2}$ is the voltage at node $2$ and is considered as the output voltage. The elements \texttt{A}, \texttt{B}, \texttt{C}, \texttt{D}, \texttt{E}, \texttt{F}, \texttt{G} and \texttt{H} of the lists \texttt{inlst\_pass\_aggres} and \texttt{outlst\_pass\_aggres} are:

\vspace*{0.2cm}

\begin{mdframed}
\begin{flushleft}
$\mathtt{\ }$\hspace{0.030cm}  \small\textup{\texttt{$\mathtt{A = R_{1a} (C_{2a} + C_{3a}) + 2 R_{2a} C_{3a} }$}} \\ \vspace*{0.04cm}
\vspace*{0.1cm}
$\mathtt{\ }$\hspace{0.030cm}  \small\textup{\texttt{$\mathtt{B = R_{2a} C_{3a} (2 R_{1a} C_{2a} + R_{1a} C_{3a} + R_{2a} C_{3a}) }$}} \\ \vspace*{0.04cm}
\vspace*{0.1cm}
$\mathtt{\ }$\hspace{0.0cm}  \small\textup{\texttt{$\mathtt{C = R_{1a} {R_{2a}}^2 C_{2a} {C_{3a}}^2}$}} \\ \vspace*{0.04cm}
$\mathtt{\ }$\hspace{0.030cm}  \small\textup{\texttt{$\mathtt{D = 2 R_{1a} (C_{2a} + C_{3a}) + 2 R_{2a} C_{3a} + R_{th} (C_{1a} + C_{2a} + C_{3a})     }$}} \\ \vspace*{0.07cm}
$\mathtt{\ }$\hspace{0.030cm}  \small\textup{\texttt{$\mathtt{E = 2 R_{1a} C_{2a} (2 R_{2a} C_{3a} + R_{th} C_{1a} + R_{th} C_{3a}) + 2 R_{th} C_{3a} (R_{1a} C_{1a} + R_{2a} C_{1a} + R_{2a} C_{2a}) +     }$}} \\ \vspace*{0.04cm}
$\mathtt{\ }$\hspace{0.8cm}  \small\textup{\texttt{$\mathtt{  ({R_{1a}}^2 + R_{1a} R_{th}) ({C_{2a}}^2 + {C_{3a}}^2) + R_{2a} {C_{3a}}^2 (R_{2a} + R_{th}) + 2 R_{1a} C_{3a} (R_{1a} C_{2a} + R_{2a} C_{3a})    }$}} \\ \vspace*{0.04cm}
\vspace*{0.1cm}
$\mathtt{\ }$\hspace{0.030cm}  \small\textup{\texttt{$\mathtt{F = 2 R_{1a} R_{2a} C_{2a} C_{3a} (R_{1a} C_{2a} + R_{1a} C_{3a} + R_{2a} C_{3a} + 2 R_{th} C_{1a}) +   }$}} \\ \vspace*{0.04cm}
\vspace*{0.1cm}
$\mathtt{\ }$\hspace{3.0cm}  \small\textup{\texttt{$\mathtt{  2 R_{1a} R_{th} C_{3a} (R_{2a} {C_{2a}}^2 + R_{2a} C_{2a} C_{3a}  + R_{1a} C_{1a} C_{2a} + R_{2a} C_{1a} C_{3a}) +    }$}} \\ \vspace*{0.04cm}
\vspace*{0.1cm}
$\mathtt{\ }$\hspace{3.0cm}  \small\textup{\texttt{$\mathtt{ {R_{1a}}^2 R_{th} C_{1a} ({C_{2a}}^2 + {C_{3a}}^2) + {R_{2a}}^2 R_{th} {C_{3a}}^2 (C_{1a} + C_{2a})  }$}} \\ \vspace*{0.04cm}
\vspace*{0.1cm}
$\mathtt{\ }$\hspace{0.030cm}  \small\textup{\texttt{$\mathtt{G = R_{1a} {R_{2a}}^2 {C_{2a}}^2 {C_{3a}}^2 (R_{1a} + R_{th}) + 2 R_{1a} R_{2a} R_{th} C_{1a} C_{2a} C_{3a} (R_{1a} C_{2a} + R_{1a} C_{3a} + R_{2a} C_{3a})  }$}} \\ \vspace*{0.04cm}
\vspace*{0.1cm}
$\mathtt{\ }$\hspace{0.030cm}  \small\textup{\texttt{$\mathtt{H = {R_{1a}}^2 {R_{2a}}^2 R_{th} C_{1a} {C_{2a}}^2 {C_{3a}}^2 }$}}
\end{flushleft}
\end{mdframed}


We verified the transfer function of the passive aggressor as follows:

\vspace*{0.2cm}

\begin{mdframed}
	\begin{flushleft}
		\begin{theorem}
			\label{THM:implem_imp_tf_model_pass_aggres}
\emph{Transfer Function Verification of Passive Aggressor} \\{\small
\textup{\texttt{$\vdash$ $\forall$ $\mathtt{R_{1a}\ R_{2a}\ R_{th}\ C_{1a}\ C_{2a}\ C_{3a}\ V_{in}\ V_2}$ s.  \\
   $\mathtt{\ }$\hspace{0.8cm} 0 < $\mathtt{R_{1a}}$ $\wedge$ 0 < $\mathtt{R_{2a}}$ $\wedge$ 0 < $\mathtt{R_{th}}$ $\wedge$ \\
 $\mathtt{\ }$\hspace{0.8cm} 0 < $\mathtt{C_{1a}}$ $\wedge$ 0 < $\mathtt{C_{2a}}$ $\wedge$ 0 < $\mathtt{C_{3a}}$ $\wedge$  \\ \vspace*{0.1cm}
$\mathtt{\ }$\hspace{0.8cm} laplace\_transform $\mathtt{V_{in}}$ s $\neq$ ${\mathtt{{\overrightarrow{\mathtt{0}}}^2}}$  $\wedge$ \\ \vspace*{0.1cm}
$\mathtt{\ }$\hspace{0.8cm} $\mathtt{{\overrightarrow{\texttt{H}}}^2}$$\mathtt{s^5}$ + $\mathtt{{\overrightarrow{\texttt{G}}}^2}$$\mathtt{s^4}$ + $\mathtt{{\overrightarrow{\texttt{F}}}^2}$$\mathtt{s^3}$ + $\mathtt{{\overrightarrow{\texttt{E}}}^2}$$\mathtt{s^2}$ + $\mathtt{{\overrightarrow{\texttt{D}}}^2}$s + $\mathtt{{\overrightarrow{\texttt{1}}}^2}$  $\neq$ ${\mathtt{{\overrightarrow{\mathtt{0}}}^2}}$  $\wedge$  \\
$\mathtt{\ }$\hspace{0.8cm} zero\_initial\_conditions $\mathtt{V_{in}\ V_2}$  $\wedge$ \\
$\mathtt{\ }$\hspace{0.8cm} ($\forall$ t.\ differentiable\_higher\_derivative $\mathtt{V_{in}\ V_2}$ t) $\wedge$  \\
   $\mathtt{\ }$\hspace{0.8cm} laplace\_exists\_higher\_deriv $\mathtt{V_{in}\ V_2}$ s $\wedge$   \\
  $\mathtt{\ }$\hspace{0.8cm} ($\forall$ t.\ pass\_aggressor\_behav\_spec $\mathtt{R_{1a}\ R_{2a}\ R_{th}\ C_{1a}\ C_{2a}\ C_{3a}\ V_{in}\ V_2}$ t)   \\
\vspace*{0.2cm}
$\mathtt{\ }$\hspace{1.5cm} $\Rightarrow$ $\mathtt{\dfrac{laplace\_transform\ V_2\ s}{laplace\_transform\ V_{in}\ s}}$ =   \\
\vspace*{0.2cm}
$\mathtt{\ }$\hspace{4.0cm} $\mathtt{\dfrac{\mathtt{{\overrightarrow{\texttt{C}}}^2} s^3 + \mathtt{{\overrightarrow{\texttt{B}}}^2} s^2 + \mathtt{{\overrightarrow{\texttt{A}}}^2} s + \mathtt{{\overrightarrow{\texttt{1}}}^2}}{\mathtt{{\overrightarrow{\texttt{H}}}^2} s^5 + \mathtt{{\overrightarrow{\texttt{G}}}^2} s^4 + \mathtt{{\overrightarrow{\texttt{F}}}^2} s^3 + \mathtt{{\overrightarrow{\texttt{E}}}^2} s^2 + \mathtt{{\overrightarrow{\texttt{D}}}^2} s + \mathtt{{\overrightarrow{\texttt{1}}}^2}}}$
			}}}
		\end{theorem}
	\end{flushleft}
\end{mdframed}

\vspace*{0.2cm}

\noindent The first eight assumptions present the design requirements for the underlying system. The next assumption models the \textit{zero initial conditions} for the voltage functions $\mathtt{V_{in}}$ and $\mathtt{V_2}$. The next two assumptions provide the differentiability and the Laplace existence condition for the higher-order derivatives of $\mathtt{V_{in}}$ and $\mathtt{V_2}$ up to the orders $3$ and $5$, respectively. The last assumption presents the behavioural specification of the passive aggressor.
Finally, the conclusion of Theorem~\ref{THM:implem_imp_tf_model_pass_aggres} presents its required transfer function.
A notable feature is that the verification of Theorem~\ref{THM:implem_imp_tf_model_pass_aggres} is done almost automatically using the automatic tactic \texttt{DIFF\_EQ\_2\_TRANS\_FUN\_TAC}, which is developed in our proposed formalization.

Next, we verified the differential equation of the passive aggressor based on its transfer function using the following HOL Light theorem:

\vspace*{0.8cm}

\begin{mdframed}
\begin{flushleft}
\begin{theorem}
\label{THM:transfer_fun_imp_diff_eq_pass_aggres}
\emph{Differential Equation Verification of Passive Aggressor} \\{\small
\textup{\texttt{$\vdash$ $\forall$ $\mathtt{R_{1a}\ R_{2a}\ R_{th}\ C_{1a}\ C_{2a}\ C_{3a}\ V_{in}\ V_2}$ r.  \\
   $\mathtt{\ }$\hspace{0.8cm} 0 < $\mathtt{R_{1a}}$ $\wedge$ 0 < $\mathtt{R_{2a}}$ $\wedge$ 0 < $\mathtt{R_{th}}$ $\wedge$ 0 < $\mathtt{C_{1a}}$ $\wedge$ 0 < $\mathtt{C_{2a}}$ $\wedge$ 0 < $\mathtt{C_{3a}}$ $\wedge$  \\
$\mathtt{\ }$\hspace{0.8cm} \big($\forall$ s.\ Re$($r$)$ $\le$ Re$($s$)$ $\Rightarrow$ laplace\_transform $\mathtt{V_{in}}$ s $\neq$ ${\mathtt{{\overrightarrow{\mathtt{0}}}^2}}$\big)  $\wedge$ \\
$\mathtt{\ }$\hspace{0.8cm} \Big($\forall$ s.\ Re$($r$)$ $\le$ Re$($s$)$ $\Rightarrow$  \\
$\mathtt{\ }$\hspace{2.5cm} $\mathtt{{\overrightarrow{\texttt{H}}}^2}$$\mathtt{s^5}$ + $\mathtt{{\overrightarrow{\texttt{G}}}^2}$$\mathtt{s^4}$ + $\mathtt{{\overrightarrow{\texttt{F}}}^2}$$\mathtt{s^3}$ + $\mathtt{{\overrightarrow{\texttt{E}}}^2}$$\mathtt{s^2}$ + $\mathtt{{\overrightarrow{\texttt{D}}}^2}$s + $\mathtt{{\overrightarrow{\texttt{1}}}^2}$  $\neq$ ${\mathtt{{\overrightarrow{\mathtt{0}}}^2}}$\Big)  $\wedge$  \\
$\mathtt{\ }$\hspace{0.8cm}    zero\_initial\_conditions $\mathtt{V_{in}}$ $\mathtt{V_2}$ $\wedge$  \\
$\mathtt{\ }$\hspace{0.8cm}  \big($\forall$ t.\ differentiable\_higher\_derivative $\mathtt{V_{in}}$ $\mathtt{V_2}$ t\big) $\wedge$  \\ \vspace*{0.1cm}
$\mathtt{\ }$\hspace{0.8cm} 0 < Re$($r$)$ $\wedge$   \\ \vspace*{0.1cm}
$\mathtt{\ }$\hspace{0.8cm} \big($\forall$ s.\ Re$($r$)$ $\le$ Re$($s$)$ $\Rightarrow$ laplace\_exists\_higher\_deriv 3 $\mathtt{V_{in}}$ s\big) $\wedge$   \\ \vspace*{0.05cm}
$\mathtt{\ }$\hspace{0.8cm}  \big($\forall$ s.\ Re$($r$)$  $\le$ Re$($s$)$  $\Rightarrow$ laplace\_exists\_higher\_deriv 5 $\mathtt{V_2}$ s\big) $\wedge$   \\ \vspace*{0.1cm}
  $\mathtt{\ }$\hspace{0.8cm}  \Bigg($\forall$ s. Re$($r$)$ $\le$ Re$($s$)$ $\Rightarrow$     \\
$\mathtt{\ }$\hspace{2.4cm} $\mathtt{
\dfrac{\texttt{laplace\_transform $\mathtt{V_2}$ s}}{\texttt{laplace\_transform $\mathtt{V_{in}}$ s}} =
}$    \\
\vspace*{0.2cm}
$\mathtt{\ }$\hspace{4.4cm} $\mathtt{\dfrac{\mathtt{{\overrightarrow{\texttt{C}}}^2} s^3 + \mathtt{{\overrightarrow{\texttt{B}}}^2} s^2 + \mathtt{{\overrightarrow{\texttt{A}}}^2} s + \mathtt{{\overrightarrow{\texttt{1}}}^2}}{\mathtt{{\overrightarrow{\texttt{H}}}^2} s^5 + \mathtt{{\overrightarrow{\texttt{G}}}^2} s^4 + \mathtt{{\overrightarrow{\texttt{F}}}^2} s^3 + \mathtt{{\overrightarrow{\texttt{E}}}^2} s^2 + \mathtt{{\overrightarrow{\texttt{D}}}^2} s + \mathtt{{\overrightarrow{\texttt{1}}}^2}}}$
  \Bigg) \\
      $\mathtt{\ }$\hspace{0.8cm}  $\ $     \\
          $\mathtt{\ }$\hspace{1.2cm}  $\Rightarrow$ \Bigg($\forall$ t. 0 $\le$ $\mathtt{\underline{t}}$ $\Rightarrow$ pass\_aggressor\_behav spec      \\
 $\mathtt{\ }$\hspace{6.3cm}  $\mathtt{R_{1a}\ R_{2a}\ R_{th}\ C_{1a}\ C_{2a}\ C_{3a}\ V_{in}\ V_2}$ t\Bigg)
}}}
\end{theorem}
\end{flushleft}
\end{mdframed}

The first ten assumptions are the same as that of Theorem~\ref{THM:implem_imp_tf_model_pass_aggres}. The next assumption ensures that the real part of the Laplace variable $\texttt{r}$ is always positive. The next two assumptions describe the differentiability condition for the functions $\mathtt{V_{in}}$ and $\mathtt{V_2}$ and their higher derivatives up to the order $3$ and $5$, respectively. The last assumption provides the transfer function of the passive aggressor. Finally, the conclusion presents the corresponding differential equation of the passive aggressor.
The verification of Theorem~\ref{THM:transfer_fun_imp_diff_eq_pass_aggres} is done almost automatically using the automatic tactic \texttt{TRANS\_FUN\_2\_DIFF\_EQ\_TAC}, which is also developed in our proposed formalization.



\subsection{Formal Analysis of Passive Victim}\label{SUBSEC:form_anal_pass_victim}

\begin{figure}[ht!]
\centering
\scalebox{0.55}
{\includegraphics[trim={2.0 0.5cm 2.0 0.5cm},clip]{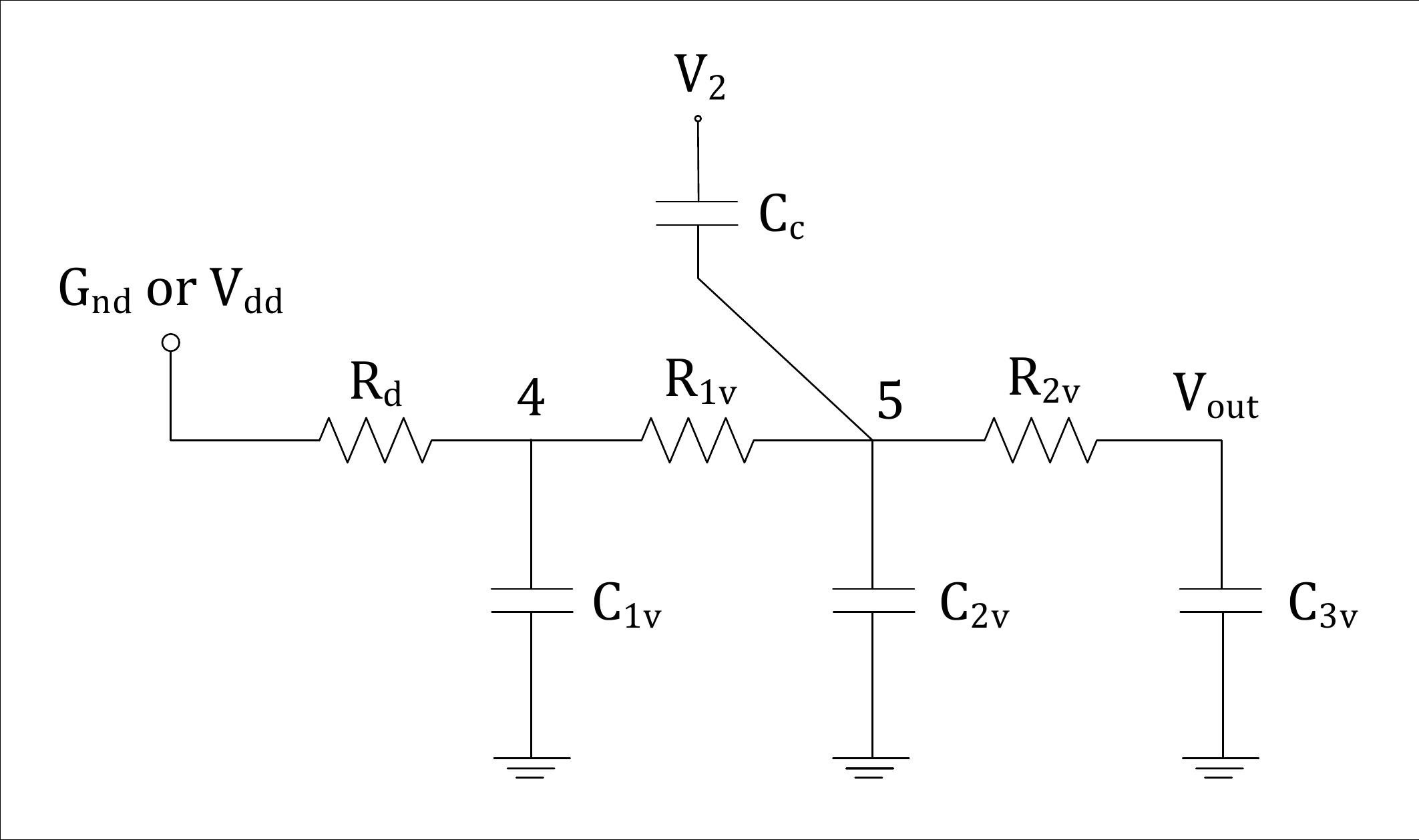}}
\caption{Passive Victim Model~\cite{sayil2007precise}}
\label{FIG:victim_model}
\end{figure}

Based on the $4$-$\pi$ interconnect circuit, Figure~\ref{FIG:victim_model} depicts the passive victim model for analyzing the crosstalk noise and delay. The resistance $R_{d}$ is the effective resistance of the victim driver~\cite{sayil2016soft,sayil2007precise}.

We model the dynamical behaviour, i.e., the modeling differential equation of the passive victim
using the $n$-order differential equation as follows:

\begin{mdframed}
\begin{flushleft}
\begin{definition}
\label{DEF:pass_victim_behav_model}
\emph{Behavioural Specification of Passive Victim} \\{
\small
\textup{\texttt{$\vdash$ $\forall$ $\mathtt{R_{1v}\ R_{2v}\ R_{d}\ C_c\ C_{1v}\ C_{3v}}$.   }}} \\
$\mathtt{\ }$\hspace*{0.8cm}  \small\textup{\texttt{inlst\_pass\_victim $\mathtt{R_{1v}\ R_{2v}\ R_{d}\ C_c\ C_{1v}\ C_{3v}}$ = [$\mathtt{{\overrightarrow{\mathtt{0}}}^2}$; $\mathtt{{\overrightarrow{\mathtt{A'}}}^2}$; $\mathtt{{\overrightarrow{\mathtt{B'}}}^2}$; $\mathtt{{\overrightarrow{\mathtt{C'}}}^2}$]}} \\
{\small
\textup{\texttt{$\vdash$ $\forall$ $\mathtt{R_{1v}\ R_{2v}\ R_{d}\ C_c\ C_{1v}\ C_{2v}\ C_{3v}}$.  }}}\\
$\mathtt{\ }$\hspace*{0.8cm}  \small\textup{\texttt{ outlst\_pass\_victim $\mathtt{R_{d}\ R_{1v}\ R_{2v}\ C_c\ C_{1v}\ C_{2v}\ C_{3v}}$ = }} \\
$\mathtt{\ }$\hspace*{6.8cm}  \small\textup{\texttt{ \big[$\mathtt{{\overrightarrow{\mathtt{1}}}^2}$; $\mathtt{{\overrightarrow{\mathtt{D'}}}^2}$; $\mathtt{{\overrightarrow{\mathtt{E'}}}^2}$; $\mathtt{{\overrightarrow{\mathtt{F'}}}^2}$; $\mathtt{{\overrightarrow{\mathtt{G'}}}^2}$\big]}} \\
{\small
\textup{\texttt{$\vdash$ $\forall$ $\mathtt{V_2\ V_{out}\ C_c\ C_{1v}\ C_{2v}\ C_{3v}\ R_{d}\ R_{1v}\ R_{2v}}$ t. \\
$\mathtt{\ }$\hspace{0.8cm}  pass\_victim\_behav\_spec $\mathtt{R_{1v}\ R_{2v}\ R_{d}\ C_c\ C_{1v}\ C_{2v}\ C_{3v}\ V_2\ V_{out}}$ t $\Leftrightarrow$ \\
$\mathtt{\ }$\hspace{0.8cm}  diff\_eq\_n\_order 4  \\
$\mathtt{\ }$\hspace{1.8cm}  (outlst\_pass\_victim $\mathtt{R_{1v}\ R_{2v}\ R_{d}\ C_c\ C_{1v}\ C_{2v}\ C_{3v}}$) $\mathtt{V_{out}}$ t = \\
$\mathtt{\ }$\hspace{0.8cm}  diff\_eq\_n\_order 3   \\
$\mathtt{\ }$\hspace{1.8cm}  (inlst\_pass\_victim $\mathtt{R_{1v}\ R_{2v}\ R_{d}\ C_c\ C_{1v}\ C_{3v}}$) $\mathtt{V_2}$ t
}}}
\end{definition}
\end{flushleft}
\end{mdframed}

\noindent where $\mathtt{V_2}$ and $\mathtt{V_{out}}$ are the input and output voltages, respectively, having data types $\mathtt{\mathds{R}^1 \rightarrow \mathds{R}^2}$. The elements $\mathtt{A'}$, $\mathtt{B'}$, $\mathtt{C'}$, $\mathtt{D'}$, $\mathtt{E'}$, $\mathtt{F'}$ and $\mathtt{G'}$ of the lists \texttt{inlst\_pass\_victim} and \texttt{outlst\_pass\_victim} are:

\begin{mdframed}
\begin{flushleft}
$\mathtt{\ }$\hspace{0.030cm}  \small\textup{\texttt{$\mathtt{A' = (R_d + R_{1v}) C_c }$}} \\
\vspace*{0.1cm}
$\mathtt{\ }$\hspace{0.030cm}  \small\textup{\texttt{$\mathtt{B' = R_{2v} C_c C_{3v} (R_d + R_{1v}) + R_d R_{1v} C_c C_{1v} }$}} \\
\vspace*{0.1cm}
$\mathtt{\ }$\hspace{0.0cm}  \small\textup{\texttt{$\mathtt{C' = R_{d} R_{1v} R_{2v} C_c C_{1v} C_{3v}}$}} \\
\vspace*{0.1cm}
$\mathtt{\ }$\hspace{0.030cm}  \small\textup{\texttt{$\mathtt{D' = R_d ( C_c + C_{1v} + C_{2v} + C_{3v}) + R_{1v} (C_c + C_{2v} + C_{3v}) + 2 R_{2v} C_{3v}   }$}} \\
\vspace*{0.1cm}
$\mathtt{\ }$\hspace{0.030cm}  \small\textup{\texttt{$\mathtt{E' = R_d R_{2v} C_{3v}  (2 C_c + 2 C_{1v} + 2 C_{2v} + C_{3v}) + R_{1v} R_{2v} C_{3v} (2 C_{2v} + 2 C_c + C_{3v}) +  }$}} \\
\vspace*{0.1cm}
$\mathtt{\ }$\hspace{6.5cm}  \small\textup{\texttt{$\mathtt{ R_d R_{1v} C_{1v} (C_c + C_{2v} + C_{3v}) + {R_{2v}}^2 {C_{3v}}^2  }$}} \\
\vspace*{0.1cm}
$\mathtt{\ }$\hspace{0.030cm}  \small\textup{\texttt{$\mathtt{F' = R_d R_{1v} R_{2v} C_{1v} C_{3v} (2 C_c + 2 C_{2v} + C_{3v}) +   }$}} \\
\vspace*{0.1cm}
$\mathtt{\ }$\hspace{4.0cm}  \small\textup{\texttt{$\mathtt{ {R_{2v}}^2 {C_{3v}}^2 \big[R_d (C_c + C_{1v} + C_{2v}) + R_{1v} (C_c + C_{2v})\big] }$}} \\
\vspace*{0.1cm}
$\mathtt{\ }$\hspace{0.030cm}  \small\textup{\texttt{$\mathtt{G' = R_d R_{1v} {R_{2v}}^2 C_{1v} {C_{3v}}^2 (C_c + C_{2v}) }$}}
\end{flushleft}
\end{mdframed}

We verified the transfer function of the passive victim as follows:

\begin{mdframed}
	\begin{flushleft}
		\begin{theorem}
			\label{THM:implem_imp_tf_model_pass_victim}
\emph{Transfer Function Verification of Passive Victim} \\{\small
\textup{\texttt{$\vdash$ $\forall$ $\mathtt{R_{1v}\ R_{2v}\ R_{d}\ C_{c}\ C_{1v}\ C_{2v}\ C_{3v}\ V_2\ V_{out}}$ s.  \\
   $\mathtt{\ }$\hspace{0.8cm} 0 < $\mathtt{R_{1v}}$ $\wedge$ 0 < $\mathtt{R_{2v}}$ $\wedge$ 0 < $\mathtt{R_d}$ $\wedge$ \\
 $\mathtt{\ }$\hspace{0.8cm} 0 < $\mathtt{C_{1v}}$ $\wedge$ 0 < $\mathtt{C_{2v}}$ $\wedge$ 0 < $\mathtt{C_{3v}}$ $\wedge$ 0 < $\mathtt{C_c}$ $\wedge$ \\ \vspace*{0.1cm}
$\mathtt{\ }$\hspace{0.8cm} laplace\_transform $\mathtt{V_2}$ s $\neq$ ${\mathtt{{\overrightarrow{\mathtt{0}}}^2}}$  $\wedge$ \\ \vspace*{0.1cm}
$\mathtt{\ }$\hspace{0.8cm} ${\mathtt{{\overrightarrow{\mathtt{G'}}}^2}}$$\mathtt{s^4}$ + ${\mathtt{{\overrightarrow{\mathtt{F'}}}^2}}$$\mathtt{s^3}$ + ${\mathtt{{\overrightarrow{\mathtt{E'}}}^2}}$$\mathtt{s^2}$ + ${\mathtt{{\overrightarrow{\mathtt{D'}}}^2}}$s + ${\mathtt{{\overrightarrow{\mathtt{1}}}^2}}$  $\neq$ ${\mathtt{{\overrightarrow{\mathtt{0}}}^2}}$  $\wedge$  \\
    $\mathtt{\ }$\hspace{0.8cm} zero\_initial\_conditions $\mathtt{V_2\ V_{out}}$  $\wedge$ \\
$\mathtt{\ }$\hspace{0.8cm} ($\forall$ t.\ differentiable\_higher\_derivative $\mathtt{V_2\ V_{out}}$ t) $\wedge$  \\
   $\mathtt{\ }$\hspace{0.8cm} laplace\_exists\_higher\_deriv $\mathtt{V_2\ V_{out}}$ s $\wedge$   \\
  $\mathtt{\ }$\hspace{0.8cm} ($\forall$ t.\ pass\_victim\_behav\_spec $\mathtt{R_{1v}\ R_{2v}\ R_d\ C_c\ C_{1v}\ C_{2v}\ C_{3v}\ V_2\ V_{out}}$ t)   \\
\vspace*{0.1cm}
$\mathtt{\ }$\hspace{1.5cm} $\Rightarrow$ $\mathtt{\dfrac{laplace\_transform\ V_{out}\ s}{laplace\_transform\ V_2\ s}}$ =       $\mathtt{\dfrac{s \Big({\mathtt{{\overrightarrow{\mathtt{C'}}}^2}} s^2 + {\mathtt{{\overrightarrow{\mathtt{B'}}}^2}} s + {\mathtt{{\overrightarrow{\mathtt{A'}}}^2}}\Big)}{{\mathtt{{\overrightarrow{\mathtt{G'}}}^2}} s^4 + {\mathtt{{\overrightarrow{\mathtt{F'}}}^2}} s^3 + {\mathtt{{\overrightarrow{\mathtt{E'}}}^2}} s^2 + {\mathtt{{\overrightarrow{\mathtt{D'}}}^2}} s + {\mathtt{{\overrightarrow{\mathtt{1}}}^2}}}}$
			}}}
		\end{theorem}
	\end{flushleft}
\end{mdframed}

\noindent The first nine assumptions present the design requirements for the underlying system. The next assumption models the \textit{zero initial conditions} for the voltage functions $\mathtt{V_2}$ and $\mathtt{V_{out}}$. The next two assumptions provide the differentiability and the Laplace existence condition for the higher-order derivatives of $\mathtt{V_2}$ and $\mathtt{V_{out}}$ up to the orders $3$ and $4$, respectively. The last assumption presents the behavioural specification of the passive victim.
Finally, the conclusion of Theorem~\ref{THM:implem_imp_tf_model_pass_victim} presents its required transfer function.

Now, we verified the differential equation of the passive victim based on its transfer function using the following HOL Light theorem:


\begin{mdframed}
\begin{flushleft}
\begin{theorem}
\label{THM:transfer_fun_imp_diff_eq_pass_victim}
\emph{Differential Equation Verification of Passive Victim} \\{\small
\textup{\texttt{$\vdash$ $\forall$ $\mathtt{V_{out}\ V_2\ R_{1v}\ R_{2v}\ R_d\ C_{1v}\ C_{2v}\ C_{3v}\ C_c}$ r.  \\
$\mathtt{\ }$\hspace{0.8cm}  0 < $\mathtt{R_{1v}}$ $\wedge$  0 < $\mathtt{R_{2v}}$ $\wedge$  0 < $\mathtt{R_d}$ $\wedge$ 0 < $\mathtt{C_c}$ $\wedge$  \\
$\mathtt{\ }$\hspace{0.8cm}  0 < $\mathtt{C_{1v}}$ $\wedge$  0 < $\mathtt{C_{2v}}$ $\wedge$  0 < $\mathtt{C_{3v}}$ $\wedge$   \\
$\mathtt{\ }$\hspace{0.8cm} \big($\forall$ s.\ Re$($r$)$ $\le$ Re$($s$)$ $\Rightarrow$ laplace\_transform $\mathtt{V_2}$ s $\neq$ ${\mathtt{{\overrightarrow{\mathtt{0}}}^2}}$\big)  $\wedge$ \\ \vspace*{0.1cm}
$\mathtt{\ }$\hspace{0.8cm} \Big($\forall$ s.\ Re$($r$)$ $\le$ Re$($s$)$ $\Rightarrow$  \\
$\mathtt{\ }$\hspace{4.0cm} ${\mathtt{{\overrightarrow{\mathtt{G'}}}^2}}$$\mathtt{s^4}$ + ${\mathtt{{\overrightarrow{\mathtt{F'}}}^2}}$$\mathtt{s^3}$ + ${\mathtt{{\overrightarrow{\mathtt{E'}}}^2}}$$\mathtt{s^2}$ + ${\mathtt{{\overrightarrow{\mathtt{D'}}}^2}}$s + ${\mathtt{{\overrightarrow{\mathtt{1}}}^2}}$  $\neq$ ${\mathtt{{\overrightarrow{\mathtt{0}}}^2}}$\Big)  $\wedge$  \\ \vspace*{0.1cm}
$\mathtt{\ }$\hspace{0.8cm}  zero\_initial\_conditions $\mathtt{V_2}$ $\mathtt{V_{out}}$ $\wedge$  \\
$\mathtt{\ }$\hspace{0.8cm}  \big($\forall$ t.\ differentiable\_higher\_derivative $\mathtt{V_2}$ $\mathtt{V_{out}}$ t\big) $\wedge$  \\ \vspace*{0.1cm}
$\mathtt{\ }$\hspace{0.8cm}  0 < Re$($r$)$ $\wedge$    \\ \vspace*{0.1cm}
$\mathtt{\ }$\hspace{0.8cm}  \big($\forall$ s.\ Re$($r$)$ $\le$ Re$($s$)$ $\Rightarrow$ laplace\_exists\_higher\_deriv 2 $\mathtt{V_2}$ s\big) $\wedge$   \\
$\mathtt{\ }$\hspace{0.8cm}  \big($\forall$ s.\ Re$($r$)$ $\le$ Re$($s$)$ $\Rightarrow$ laplace\_exists\_higher\_deriv 4 $\mathtt{V_{out}}$ s\big) $\wedge$   \\ \vspace*{0.1cm}
  $\mathtt{\ }$\hspace{0.8cm}  \Bigg($\forall$ s. Re$($r$)$ $\le$ Re$($s$)$ $\Rightarrow$     \\
$\mathtt{\ }$\hspace{1.5cm} $\mathtt{\dfrac{laplace\_transform\ V_{out}\ s}{laplace\_transform\ V_2\ s}}$ =       $\mathtt{\dfrac{s \Big({\mathtt{{\overrightarrow{\mathtt{C'}}}^2}} s^2 + {\mathtt{{\overrightarrow{\mathtt{B'}}}^2}} s + {\mathtt{{\overrightarrow{\mathtt{A'}}}^2}}\Big)}{{\mathtt{{\overrightarrow{\mathtt{G'}}}^2}} s^4 + {\mathtt{{\overrightarrow{\mathtt{F'}}}^2}} s^3 + {\mathtt{{\overrightarrow{\mathtt{E'}}}^2}} s^2 + {\mathtt{{\overrightarrow{\mathtt{D'}}}^2}} s + {\mathtt{{\overrightarrow{\mathtt{1}}}^2}}}}$
  \Bigg) \\
      $\mathtt{\ }$\hspace{0.2cm}  $\ $     \\
          $\mathtt{\ }$\hspace{2.2cm}  $\Rightarrow$ \Bigg($\forall$ t. 0 $\le$ $\mathtt{\underline{t}}$ $\Rightarrow$ pass\_aggressor\_behav\_spec      \\
 $\mathtt{\ }$\hspace{6.2cm}  $\mathtt{V_2\ V_{out}\ C_c\ C_{1v}\ C_{2v}\ C_{3v}\ R_{1v}\ R_{2v}\ R_d}$ t\Bigg)
}}}
\end{theorem}
\end{flushleft}
\end{mdframed}

The first eleven assumptions of the above theorem are the same as that of Theorem~\ref{THM:implem_imp_tf_model_pass_victim}. The next assumption ensures that the real part of the Laplace variable $\texttt{r}$ is always positive.
The next two assumptions model the existence condition of the Laplace transform for the functions $\mathtt{V_2}$, $\mathtt{V_{out}}$ and their higher derivatives up to the order $3$ and $4$, respectively.
The last assumption provides the transfer function of the passive victim. Finally, the conclusion presents its corresponding differential equation.
The verification of Theorem~\ref{THM:transfer_fun_imp_diff_eq_pass_victim} is done almost automatically using the automatic tactic \texttt{TRANS\_FUN\_2\_DIFF\_EQ\_TAC}.

Finally, the transfer function of the overall system is represented by the following mathematical equation.

\begin{equation}\label{EQ:tf_overall_crosstalk_model}
\dfrac{V_{out} (s)}{V_{in} (s)} = \dfrac{V_{out} (s)}{V_2 (s)} \times \dfrac{V_2 (s)}{V_{in} (s)}
\end{equation}

\noindent We also verified the above transfer function and its corresponding differential equation based on our formalization and the details about their verification can be found in the proof script~\cite{adnan16lerchsthm}.
The formal analysis of the $4$-$\pi$ soft error crosstalk model is done almost automatically, thanks to our automatic tactics \texttt{DIFF\_EQ\_2\_TRANS\_FUN\_} \texttt{TAC} and \texttt{TRANS\_FUN\_2\_DIFF\_EQ\_TAC}, which are developed as part of the reported work and illustrate the usefulness of our proposed formalization of the Laplace transform in the analysis of safety-critical systems.
The distinguishing feature of Theorems~\ref{THM:transfer_fun_imp_diff_eq_pass_aggres} and~\ref{THM:transfer_fun_imp_diff_eq_pass_victim} is the relationship between the differential equation, which is expressed in the time domain, and the corresponding transfer function, which is expressed in frequency domain. However, Theorems~\ref{THM:implem_imp_tf_model_pass_aggres} and~\ref{THM:implem_imp_tf_model_pass_victim} verified using our earlier formalization~\cite{taqdees2013formalization,rashid2017formal} are completely based on the frequency domain and no relation with the commonly used differential equation is established. The formally verified Lerch's theorem allowed us to transform the problem of solving a differential equation in time domain to a problem of solving a linear equation in the frequency domain. This linear equation can be solved to determine constraints on the values of the components to ensure a low-power and energy efficient designing of the ICs. Moreover, all the verified theorems are of generic nature, i.e, all the variables and functions are universally quantified and can thus be specialized to any particular value for the analysis of a system. Similarly, the high expressiveness of the higher-order logic enabled us to model the dynamical behaviour of the system, i.e., the differential equation in its true form and to perform its corresponding analysis.



\section{Conclusions}\label{SEC:Conclusions}

This paper presents a formalization of Lerch's theorem using the HOL Light theorem prover. This result extends our formalization of the Laplace transform, which includes the formal definition of the Laplace transform and verification of its various classical properties such as linearity, frequency shifting, differentiation and integration in time domain, time shifting, time scaling, modulation and the Laplace transform of a $n$-order differential equation. Lerch's theorem describes the uniqueness of the Laplace transform and thus can be used to find solutions of linear differential equations in the time domain, which was not possible with our earlier formalization of the Laplace transform. We used our proposed formalization for formally analyzing a $4$-$\pi$ soft error crosstalk model for the nanometer technologies.

In the future, we aim to formally verify the uniqueness of the Fourier transform using the reported formalization of Lerch's theorem. The region of integration for the case of Fourier transform is $(-\infty,\infty)$~\cite{rashid2016formalization}, whereas, the one in the case of Laplace is from $[0,\infty)$. We can split region of the integration for the integral of the Fourier transform into two sub-intervals: $(-\infty,0]$ and $[0,\infty)$. The uniqueness of the first integral can be directly handled by Lerch's theorem, whereas, for the case of $(-\infty,0]$, the integral can be first reflected and then the formally verified Lerch's theorem can be used to verify its uniqueness as well. Another future direction is to use this formalization in our project on system biology~\cite{rashid2017formalreasoning}, for finding the analytical solutions of the differential equation based reaction kinetic models of the biological systems.


  \bibliographystyle{plain}
  \bibliography{bibliotex}

\end{document}